\documentclass[usenatbib]{mn2e}
\usepackage{natbibmnfix,graphicx,times}

\newcommand{\cmsq}{\mbox{ cm$^2$}}

\newcommand{\kel}{\mbox{ K}}

\newcommand{\Mpc}{\mbox{ Mpc}}
\newcommand{\Mpcden}{\mbox{ Mpc$^{-3}$}}

\newcommand{\msun}{\mbox{ M$_\odot$}}
\newcommand{\sfr}{\mbox{ M$_\odot$ yr$^{-1}$}}

\newcommand{\hunits}{\mbox{ km s$^{-1}$ Mpc$^{-1}$}}
\newcommand{\kms}{\mbox{ km s$^{-1}$}}
\newcommand{\bq}{\begin{equation}}
\newcommand{\eq}{\end{equation}}
\newcommand{\bqa}{\begin{eqnarray}}
\newcommand{\eqa}{\end{eqnarray}}

\newcommand{\bxh}{\bar{x}_H}
\newcommand{\xh}{x_H}
\newcommand{\Qb}{\bar{Q}}

\title[Ly$\alpha$ Absorption During Reionization]{ Constraining the
  Topology of Reionization Through Ly$\alpha$ Absorption}
 
\author[S.~Furlanetto, L.~Hernquist, \&
M. Zaldarriaga]{S.~R.~Furlanetto,$^1$\thanks{Email:
sfurlane@tapir.caltech.edu}, L.~Hernquist,$^2$ and
M.~Zaldarriaga$^{2,3}$ \\
$^1$Division of Physics, Mathematics, \& Astronomy;
California Institute of Technology; Mail Code 130-33; Pasadena, CA
91125, USA \\
$^2$Harvard-Smithsonian Center for Astrophysics, 60
Garden St., Cambridge, MA 02138, USA \\
$^3$Jefferson Laboratory of Physics, Harvard University, 
Cambridge, MA 02138, USA}
 
\begin{document}
 
\maketitle
 






\begin{abstract}
The reionization of hydrogen in the intergalactic medium (IGM) is a
crucial landmark in the history of the universe, but the processes
through which it occurs remain mysterious.  In particular, recent
numerical and analytic work suggest that reionization by stellar
sources is driven by large-scale density fluctuations and must be
inhomogeneous on scales of many comoving Mpc.  We examine the
prospects for constraining the topology of neutral and ionized gas
through Ly$\alpha$ absorption of high-redshift sources.  One method is
to search for gaps in the Gunn-Peterson absorption troughs of luminous
sources.  These could occur if the line of sight passes sufficiently
close to the center of a large HII region.  In contrast to previous
work, we find a non-negligible (though still small) probability of
observing such a gap before reionization is complete.  In our model
the transmission spike at $z=6.08$ in the spectrum of SDSS J1148+5251
does not necessarily require overlap to have been completed at an
earlier epoch.  We also examine the IGM damping wing absorption of
the Ly$\alpha$ emission lines of star-forming galaxies.  Because most
galaxies sit inside of large HII regions, we find that the severity
of absorption is significantly smaller than previously thought and
decoupled from the properties of the observed galaxy.  While this
limits our ability to constrain the mean neutral fraction of the IGM
from observations of individual galaxies, it presents the exciting
possibility of measuring the size distribution and evolution of the
ionized bubbles by examining the distribution of damping wing optical
depths in a large sample of galaxies.

\end{abstract}
  

\section{Introduction}

The reionization of the intergalactic medium (IGM) constitutes a
milestone in the history of the universe, because it marks the epoch
at which (radiative) feedback from the first generations of luminous
objects transformed the universe on the largest scales.  It therefore
offers a fascinating probe of both the IGM and these first luminous
sources, and astronomers have focussed a great deal of attention on
understanding the transition.  Intriguingly, three independent
observational techniques offer constraints that appear (at first
sight) mutually contradictory.  One comes from spectra of
high-redshift quasars identified through the Sloan Digital Sky Survey
(SDSS).\footnote{See http://www.sdss.org/ for more information on the
SDSS.}  All four quasars with $z > 6.2$ show complete \citet{gp}
absorption over substantial path lengths in their spectrum
\citep{becker01,fan03,white03,fan04}; this implies a mean neutral
fraction $\bxh \ga 10^{-3}$ and a rapid change in the ionizing
background at this epoch (\citealt{fan}; but see \citealt{songaila04}
for a different interpretation).  Both of these properties indicate
that the tail end of reionization occurs at $z\sim6$ (see also
\citealt{wyithe04-prox}).  A second constraint comes from observations
of the cosmic microwave background (CMB).  Primordial anisotropies are
washed out by free electrons after reionization, but the same process
generates a large-scale polarization signal \citep{zal97}.  This
provides an integral constraint on the reionization history; recent
measurements imply that $\bxh$ must have been small at $z \ga 14$
\citep{kogut03,spergel03}.  Finally, the relatively high temperature
of the Ly$\alpha$ forest at $z \sim 2$--$4$ suggests an order unity
change in $\bxh$ at $z \la 10$ \citep{theuns02-reion,hui03}, although
this conclusion is weakened by uncertainties about HeII
reionization (e.g., \citealt{sokasian02}).

These three observations imply that the reionization history must have
been complex and extended.  Such a history is inconsistent with
theoretical expectations for ``normal'' galaxies, which cause a simple
transition with a smoothly increasing ionized fraction (e.g.,
\citealt{barkana01}, and references therein).  The observations appear
to indicate qualitative evolution in the properties of the ionizing
sources as well as the crucial importance of feedback mechanisms
\citep{wyithe03,cen03,haiman03,sokasian03b}.  Unfortunately,
extracting more detailed information about reionization appears
difficult.  One strategy is ``21 cm tomography'' of the high-redshift
IGM (e.g., \citealt{scott,mmr,zald04,furl04b}), in which one maps the
distribution of neutral hydrogen on large scales through its
redshifted hyperfine transition.  However, this technique requires new
instruments and will likely not be possible for several years.
Detailed analyses of the CMB may also provide more information
\citep{holder03,santos03} but promise to be similarly difficult.

Here we consider the study of reionization through surveys
of high-redshift galaxies and quasars.  Such observations obviously
help by measuring the abundance and properties of luminous sources
(e.g., \citealt{barton04}).  Beyond this, however, Ly$\alpha$
absorption by neutral hydrogen in the IGM can have powerful effects on
the spectra of these sources \citep{gp,miresc98}, so detailed
observations of their properties can also be used to constrain the IGM
itself.  The first possibility is to examine the Ly$\alpha$ emission
lines of high-redshift galaxies.  Narrowband searches for such lines
are, in fact, one of the most efficient techniques to identify these
sources.  Surveys currently extend to $z \sim 6$--$7$ (e.g.,
\citealt{hue02,kodaira03,rhoads04,stanway04,santos04-obs}) and will
likely reach higher redshifts in the coming years.  Because the
Ly$\alpha$ line is subject to strong absorption from the IGM, its
characteristics can help to constrain the properties of the
surrounding gas.  One factor is that the damping wing absorption
should make strong Ly$\alpha$ emitters rarer before reionization,
although \citet{haiman02} showed that the ionized bubble around
luminous galaxies will not necessarily completely destroy the line
(see also \citealt{madau00} for a related discussion in the context of
quasars).  \citet{santos04} made a thorough study of how local
properties of the galaxy (such as winds and infall) would affect such
inferences; these properties can substantially affect the Ly$\alpha$
absorption, but because they do not depend on the ionization state of
the IGM they should not strongly affect how the relative absorption
changes through reionization.  Thus there remains the exciting
possibility of constraining the IGM's ionization state through
observations of high-redshift galaxies.  For example, the recent
detection of a possible $z=10$ galaxy by \citet{pello04} triggered a
flurry of theoretical efforts to constrain the ionization state of the
IGM and the ionizing luminosity of the galaxy through its apparent
absorption \citep{loeb04,ricotti04,gnedin04,cen04}.  (Note that,
although this object was not selected by its Ly$\alpha$ line, its
spectrum apparently nevertheless shows one.)

However, none of these theoretical studies have explicitly considered
the possibility that neighboring galaxies can alter the local
ionization state.  Because high-redshift galaxies are so strongly
clustered, any source is likely to have a number of nearby neighbors
whose own HII regions may overlap with that of the detected galaxy
even well before reionization is complete.  In this case, the
absorption could be much less than predicted by naive models in which
each galaxy is isolated.  \citet{cen03-qso} suggested that this may be
important near luminous quasars, but such objects are rare before
reionization and are certainly not representative of the large-scale
environment \citep{fan04}.  \citet[hereafter FZH04]{furl04a}, guided
by simulations \citep{sokasian03a,ciardi03-sim} and analytic work
\citep{barkana03}, developed a simple model to describe how
large-scale density fluctuations drive the growth of HII regions.
The model reproduces the qualitative features of numerical simulations
of reionization, which contain a relatively small number of large HII
regions (with sizes of several comoving Mpc) encompassing many
ionizing sources.  These create a large-scale inhomogeneous pattern of
neutral and ionized gas that evolves throughout reionization.  Most
luminous sources sit inside regions that are ionized relatively early,
so they will suffer correspondingly less absorption from the IGM.
\citet{gnedin04} used numerical simulations to examine how the
evolving bubble pattern affects the absorption suffered by individual
galaxies, although they were able to consider only one particular
reionization history and were limited by the finite size of their
simulation box.  In this paper, we will use the model of FZH04 to
quantify the amount of Ly$\alpha$ absorption expected in different
stages of reionization.  We will argue that the large HII regions
cannot be neglected when interpreting Ly$\alpha$ absorption and make
it difficult to infer the mean neutral fraction from observations of a
small number of galaxies.  However, these lines do present the
intriguing possibility of allowing us to measure the size distribution
of HII regions throughout reionization and thus to constrain the
inhomogeneity of the process.\footnote{As we were completing this
work, we became aware of a similar effort to describe the HII regions
around Ly$\alpha$-line galaxies \citep{wyithe04-lya}.  We note that
both groups reach qualitatively similar results, although each include
different aspects of the physics and utilise entirely independent
methods.}

Another technique is to study the absorption spectra of luminous
quasars.  At lower redshifts, these yield detailed information about
the density structure of the IGM through the Ly$\alpha$ forest.
Unfortunately, the depth of the Gunn-Peterson trough, which is
$\tau_{\rm GP} \sim 6.5 \times 10^5 [(1+z)/10]^{3/2}$ for a fully
neutral medium, makes it more difficult to extract information as
reionization is approached.  Before overlap, any transmission in the
trough requires passing close to a bright ionizing source; it is
particularly difficult because of the broad Ly$\alpha$ damping wing
from the surrounding IGM.  Nevertheless, the spectrum of SDSS
1148+5251 contains a transmission spike at $z=6.08$ in both the
Ly$\alpha$ and Ly$\beta$ troughs with an apparent $\tau = 2.5$
\citep{white03}.  \citet{miresc98} argued that evading the damping
wing constraint required ionized bubbles with sizes $\ga 1$ physical
Mpc and suggested that such regions were only possible around bright
quasars.  \citet{barkana02-bub} showed explicitly that, if each galaxy
is considered in isolation, transmission gaps in the Gunn-Peterson
trough should be extremely rare until reionization is complete.
However, these studies did not include the biasing of sources and the
correspondingly large HII regions.  Here we use the model of FZH04 to
compute the probability of seeing a transmission gap in the
Gunn-Peterson trough.  We show that the large ionized bubbles make the
damping wing absorption considerably less significant, so there is a
small (but non-negligible) possibility of observing such a gap even
before reionization is complete.  We thus find that the transmission
gap in the spectrum of SDSS J1148+5251 does not require overlap to be
complete by $z=6.08$.

In \S \ref{model}, we briefly review the model for reionization
originally presented in FZH04.  We then consider its implications for
surveys of Ly$\alpha$ emitting galaxies in \S \ref{lyaemit} and for
quasar absorption spectra in \S \ref{gap}.  We conclude in \S
\ref{disc}.  In our numerical calculations, we will assume a
$\Lambda$CDM cosmology with $\Omega_m=0.3$, $\Omega_\Lambda=0.7$,
$\Omega_b=0.046$, $H=100 h \hunits$ (with $h=0.7$), $n=1$, and
$\sigma_8=0.9$, consistent with the most recent measurements
\citep{spergel03}.

\section{A Model for Reionization}
\label{model}

Recent numerical simulations (e.g., \citealt{sokasian03}) show that
reionization proceeds ``inside-out'' from high density clusters of
sources to voids, at least when the sources resemble star-forming
galaxies (e.g., \citealt{springel03}).  We therefore associate HII
regions with large-scale overdensities.  We assume that a galaxy of
mass $m_{\rm gal}$ can ionize a mass $\zeta m_{\rm gal}$, where
$\zeta$ is a constant that depends on the efficiency of ionizing
photon production, the escape fraction of these photons from the host
galaxy, the star formation efficiency, and the mean number of
recombinations.  Values of $\zeta \sim 10$--$40$ are reasonable for
normal star formation, but very massive stars can increase the
efficiency by an order of magnitude \citep{bromm-vms}.  The criterion
for a region to be ionized by the galaxies contained inside it is then
$f_{\rm coll} > \zeta^{-1}$, where $f_{\rm coll}$ is the fraction of
mass bound to halos above some $m_{\rm min}$.  We will assume
that this minimum mass corresponds to a virial temperature of $10^4
\kel$, at which point hydrogen line cooling becomes efficient.  In the
extended Press-Schechter model \citep{lacey}, this places a condition
on the mean overdensity within a region of mass $m$,
\bq 
\delta_m \ge \delta_x(m,z) \equiv
\delta_c(z) - \sqrt{2} K(\zeta) [\sigma^2_{\rm min} -
\sigma^2(m)]^{1/2},
\label{eq:deltax}
\eq
where $K(\zeta) = {\rm erf}^{-1}(1 - \zeta^{-1})$, 
$\sigma^2(m)$ is the variance of density fluctuations on the
scale $m$, $\sigma^2_{\rm min}=\sigma^2(m_{\rm min})$, and $\delta_c(z)$
is the critical density for collapse.

\begin{figure}
\begin{center}
\resizebox{8cm}{!}{\includegraphics{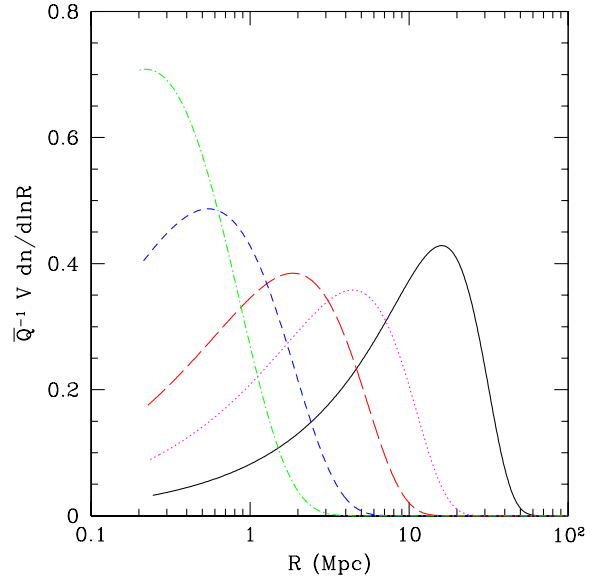}}\\%
\end{center}
\caption{ The bubble size distribution $\Qb^{-1} V dn/d \ln R$ at
  several different redshifts in our model, assuming $\zeta=40$ (note
  that $R$ is the comoving size).  Dot-dashed, short-dashed,
  long-dashed, dotted, and solid lines are for $z=18,\,16,\,14,\,13$,
  and $12$, respectively. These have $\bar{Q} = 0.037,0.11,0.3,0.5$,
  and $0.74$.}
\label{fig:dndr}
\end{figure}

FZH04 showed how to construct the mass function of HII regions from
$\delta_x$ in an analogous way to the halo mass function
\citep{press,bond91}.  The barrier in equation (\ref{eq:deltax}) is
well approximated by a linear function in $\sigma^2$, $\delta_x
\approx B(m,z) \equiv B_0 + B_1 \sigma^2(m)$. In that case the mass
function of ionized bubbles has an analytic expression
\citep{sheth98}:
\bqa
m \, n(m) \, dm & = & \sqrt{\frac{2}{\pi}} \ \frac{\bar{\rho}}{m} \ \left|
  \frac{d \ln \sigma}{d \ln m} \right| \ \frac{B_0}{\sigma(m)} 
\nonumber \\ 
& & \times \ \exp
  \left[ - \frac{B^2(m,z)}{2 \sigma^2(m)} \right] \ dm,
\label{eq:dndm}
\eqa
where $\bar\rho$ is the mean density of the universe.  Equation
(\ref{eq:dndm}) gives the comoving number density of HII regions with
masses in the range $(m,m+dm)$.  Figure \ref{fig:dndr} shows the
resulting size distributions for $\zeta=40$.  The dot-dashed,
short-dashed, long-dashed, dotted, and solid curves correspond to
$z=18,\,16,\,14,\,13$, and $12$, respectively.  We have normalized
each curve by the fraction of space $\Qb$ filled by the bubbles at the
specified redshift (i.e., $\Qb=1-\bxh$, where $\bxh$ is the mean
neutral fraction).  The curves begin at the radius corresponding to an
HII region around a completely isolated galaxy of mass $m_{\rm min}$.
Figure \ref{fig:dndr} shows that the characteristic bubble size is
much larger than this: when $\bxh \sim 0.7$ ($0.25$), the typical size
is $\sim 2$ ($16$) comoving Mpc.  Although we have shown the size
distribution for a particular choice of $\zeta$, the characteristic
scale of ionized bubbles depends primarily on $\bxh$ and is nearly
independent of $\zeta$ and $m_{\rm min}$; to first order, these
quantities simply modify the time evolution of $\bxh$.  The crucial
difference between this formula and the standard Press-Schechter mass
function arises from the fact that the barrier $\delta_x$ is a
(decreasing) function of $m$. The barrier is thus more difficult to
cross as one approaches smaller scales, which imprints a
characteristic size on the bubbles.  In contrast, the barrier used in
constructing the halo mass function, $\delta_c(z)$, is independent of
mass, which yields the usual power law behavior at small masses.

We refer the reader to FZH04 for a more in-depth discussion of this
formalism.
 
\section{Ly$\alpha$ Emitters}
\label{lyaemit}

We will first consider the how the topology of HII regions
affects galaxies with strong Ly$\alpha$ emission lines.
The essential idea is that absorption from the surrounding neutral gas
will modify the observed line; in the simplest approximation,
Ly$\alpha$ lines will be much harder to see as $\bxh$ increases
because of the increased absorption.  In \S \ref{s:prad} we calculate
the probability for a galaxy of a given mass to sit inside an ionized
bubble of a given size.  We will then describe in \S \ref{s:emitobs}
how the resulting distribution affects the prospects for surveys of
Ly$\alpha$ emitters.

\subsection{Galaxies and Their HII Regions}
\label{s:prad}

We wish to know the probability $p(m_b|m_h)$ that a halo of mass $m_h$
sits inside of an HII region with mass $m_b$ (corresponding to a
radius $R_b$).\footnote{Note that our formalism is Lagrangian.  When
converting from $m_b$ to $R_b$, we will assume that the HII regions
have the mean density.  At these early epochs and on large scales,
this is a reasonable approximation.}  One of the advantages of using
the excursion set formalism to model both the HII regions and the
dark matter halos is that it is straightforward to connect the two
objects; the argument is similar to the treatment of non-gaussianity
in \S5 of FZH04 and to the derivation of the conditional halo mass
function \citep{lacey}.  We wish to compute the probability that a
trajectory crossing the halo mass function barrier $\delta_c$ at
$\sigma^2_h$ has previously crossed the ionization barrier $\delta_x$
(or $B$ in our approximation) at $\sigma^2_b>\sigma^2_h$.  (We will
use the subscript ``b'' to refer to an ionized bubble and ``h'' to
refer to the galaxy halo.)  The distribution of first-crossings above
$\delta_c$, $p(\delta^f_h=\delta_c)$ (here the ``f'' superscript
indicates a first-crossing) corresponds to the halo mass function
\citep{press,bond91}.  Similarly, the distribution of first-crossings
above $B$, $p(\delta^f_b=B)$, can be found directly from equation
(\ref{eq:dndm}).  Moreover, given that a particle trajectory crosses
$B$ at $\sigma^2_b$ (i.e., that this particle is part of an HII
region of mass $m_b$), we can construct the conditional halo mass
function in the usual way by translating the origin to
$[\sigma^2_b,\delta=B(m_b)]$ \citep{lacey}.  We call this conditional
probability $p(\delta^f_h=\delta_c|\delta^f_b=B)$.  Then we use Bayes'
theorem to find $p(\delta^f_b=B|\delta^f_h=\delta_c)$, the probability
that a trajectory is part of an HII region of mass $m_b$ given that
it is part of a collapsed halo with mass $m_h$:
\bqa
p(\delta^f_b=B|\delta^f_h=\delta_c) p(\delta^f_h=\delta_c) d \delta_b
\ = \nonumber \\
\qquad p(\delta^f_h=\delta_c|\delta^f_b=B) p(\delta^f_b=B) d \delta_b.
\label{eq:bayes}
\eqa
The last three terms are all known, so the conditional
probability is straightforward to compute.  We can obtain the desired
result by simply transforming to mass units:
\bqa
p(m_b|m_h) d m_b & = & \sqrt{\frac{2}{\pi}} \ \frac{B_0 [\delta_c -
    B(m_b)]}{\delta_c} \ \left| \frac{d \ln \sigma_b}{d \ln
    m_b} \right| \nonumber \\
 & & \times \exp \left\{ -\frac{[\sigma_b^2 \delta_c - \sigma_h^2 B(m_b)]^2}{2
   \sigma_b^2 \sigma_h^2 (\sigma_h^2 - \sigma_b^2)} \right\} \nonumber
\\ 
& & \times \frac{\sigma_h^3}{\sigma_b
  (\sigma_h^2-\sigma_b^2)^{3/2}} \frac{dm_b}{m_b},
\label{eq:probmbub}
\eqa
where $B_0$ is defined above equation (\ref{eq:dndm}) and is found
from $B_0 \equiv \delta_x(m=\infty)$.  An implicit assumption behind
equation (\ref{eq:probmbub}) is that the density distribution
$p(\delta)$ on the relevant mass scales is gaussian.  This
approximation breaks down on sufficiently small scales, and we expect
that early in reionization (when the HII regions are small) our
formalism will be less accurate. (Note that in this regime our
linear approximation for $\delta_x$ is also only approximate.)

\begin{figure}
\begin{center}
\resizebox{8cm}{!}{\includegraphics{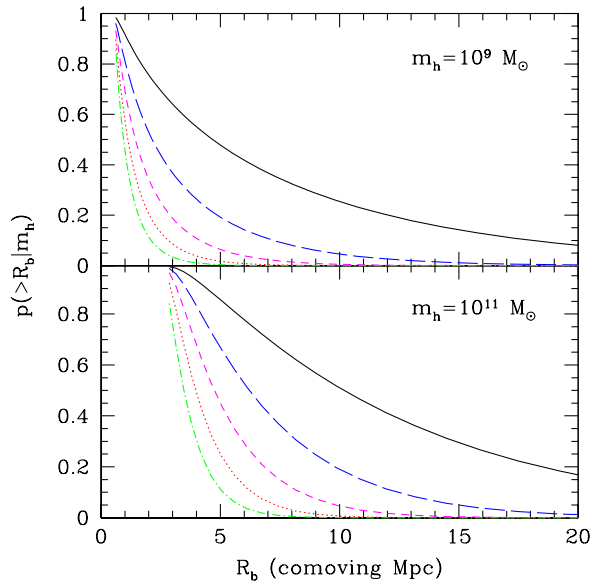}}\\%
\end{center}
\caption{The probability that halos with $m_h=10^9$ and $10^{11}
  \msun$ are embedded in an HII region above a given comoving radius
  in the $\zeta=40$ model (top and bottom panels, respectively).  The
  curves correspond to: $z=12$ ($\bxh=0.26$, solid), $z=13$
  ($\bxh=0.52$, long-dashed), $z=14$ ($\bxh=0.7$, short-dashed),
  $z=15$ ($\bxh=0.81$, dotted), and $z=16$ ($\bxh=0.89$,
  dot-dashed). }
\label{fig:raddamp}
\end{figure}

We show the cumulative probabilities that galaxies of mass $m_h=10^9
\msun$ and $m_h=10^{11} \msun$ are in bubbles with radius above some
value in Figure \ref{fig:raddamp} for the choice $\zeta=40$.  (This
choice leads to reionization at $z\sim11$.)  Each curve begins at the
minimum bubble size to contain such an ionizing galaxy, $\zeta m_h$
(i.e., the radius of the Str{\" o}mgren sphere if the galaxy had no
neighbors).  We see that including neighboring galaxies makes a
dramatic difference to the host HII regions of these galaxies.  In
both cases, nearly half are in regions at least twice the radius of
the naive estimate by the time $\bxh \sim 0.5$, and there is a
significant tail extending to much larger bubbles even earlier.  This
is a good illustration of one of the principal conclusions of FZH04:
the distribution of bubble sizes is determined not by the
characteristics of individual galaxies but instead by the large-scale
density field.  The tail at large $R_b$ is more significant for
smaller galaxies, because the most massive halos are less likely to
have neighbors of comparable or greater mass so contribute relatively
more to the local ionizing field.  Also, note that the distribution
clearly evolves rapidly throughout reionization, with the fastest
evolution occurring in the regime $\bxh \la 0.5$.

Figure \ref{fig:raddamp} reveals a minor inconsistency of our
approach: the total probability for a galaxy to be in an HII region
with $m_b>\zeta m_h$ is not equal to unity, as required by our
fundamental assumption (see \S \ref{model}).  Unfortunately, there is
nothing in the excursion set result (\ref{eq:probmbub}) to guarantee
this.  For example, imagine a trajectory that maintains $\delta \sim
0$ until $m \sim m_h$ and then rises steeply to pass through
$(\sigma^2_h,\delta_c)$.  Such a trajectory is (by construction) part
of a halo of mass $m_h$, but our formalism would assign it to an
ionized bubble of mass $m_b \sim m_h$.  Essentially this trajectory
corresponds to an isolated halo that lacks any neighbors.
Fortunately, these steep trajectories are rare, especially once the
characteristic size of HII regions exceeds $\zeta m_h$ (which happens
relatively early for most galaxy masses).  We note that in the
\citet{lacey} merger tree picture, these objects are halos that
undergo no major mergers in the future, so they must be rare in a
hierarchical structure formation scenario.  The discrepancy is worst
early in reionization, when $\delta_x$ is large and therefore more
difficult to cross.  Figure \ref{fig:raddamp} shows that, even when
$\bxh \sim 0.9$, only $\sim 15\%$ of halos are assigned to ionized
bubbles whose nominal sizes are smaller than $\zeta m_h$.  As
expected, the fraction of these trajectories decreases rapidly and
they can essentially be ignored once $\bxh \la 0.7$.  The discrepancy
is also slightly smaller for massive galaxies because
$\delta_c-\delta_x$ is an increasing function of mass (i.e., a
trajectory must rise more steeply in order to be assigned to a small
HII region).  Physically, this is because more massive galaxies are
also more highly biased.

We have shown $p(>R_b|m_h)$ for only two masses in one particular model
of reionization.  Although the bubble size distribution at a fixed
$\bxh$ is nearly independent of $\zeta$, the properties of the
ionizing galaxies clearly do change between different reionization
scenarios.  Most importantly, the characteristic mass scale of the
ionizing galaxies decreases if reionization occurs earlier.
Qualitatively, the distributions shown in Figure \ref{fig:raddamp}
would correspond to \emph{smaller} galaxies if $\zeta > 40$ and
\emph{larger} galaxies if $\zeta<40$.  Note that the two masses we
have selected are both relatively rare at $z \sim 12$--$16$; galaxies
closer to the nonlinear mass scale would experience even larger
relative effects from the HII regions of their neighbors.

\subsection{Observable Consequences}
\label{s:emitobs}

In principle, one way to constrain the ionization state of the IGM is
to measure the amount of absorption from neutral gas outside of a
galaxy's HII region.  The Hubble flow velocity offset between the
galaxy and neutral gas implies that the Ly$\alpha$ photons redshift
out of resonance by the time they encounter neutral gas.  The relevant
cross-section is therefore the ``damping wing'' of the line profile.
Assuming that the galaxy is at $z_{\rm gal}$, with neutral gas between
$z_0$ and $z_n$, the optical depth $\tau_{\rm damp}$ at $\lambda_{\rm
obs} = \lambda_\alpha (1 + z_{\rm gal})$ is \citep{miresc98}
\bqa
\tau_{\rm damp} & = & 6.43 \times 10^{-9} \xh \tau_{\rm GP}(z_{\rm gal})
\nonumber \\
& & \times \ 
\left[ I \left(\frac{1+z_n}{1+z_{\rm gal}} \right) - I
  \left(\frac{1+z_0}{1+z_{\rm gal}} \right) \right],
\label{eq:taudamp}
\eqa
where $\lambda_\alpha=1215.67$ \AA, $\tau_{\rm GP}(z_{\rm gal}) \sim 
6.5 \times 10^5 [(1+z_{\rm gal})/10]^{3/2}$  is the 
Ly$\alpha$ optical depth for a fully neutral IGM \citep{gp}, $\xh$ is
the neutral fraction outside of the HII region, and
\bqa
I(x) & = & \frac{x^{9/2}}{1-x} + \frac{9}{7} x^{7/2} + \frac{9}{5} x^{5/2}
+ 3 x^{3/2} + 9 x^{1/2} \nonumber \\ & & - \frac{9}{2} \ln \left(
\frac{1+x^{1/2}}{1-x^{1/2}} \right).
\label{eq:Idefn}
\eqa
Here we have assumed that $z_0<z_n<z_{\rm gal}$ and made the high
redshift approximation $H(z) \propto (1+z)^{3/2}$.  

The distribution of bubble sizes given in equation (\ref{eq:probmbub})
can strongly affect the damping wing absorption of a galaxy's
Ly$\alpha$ line by pushing the neutral gas much farther from the
galaxy than one would naively expect.  By decreasing $z_n$, we
increase the velocity separation between the emission line and the
neutral gas, so that the galaxy line suffers less absorption.
Neglecting peculiar velocities, the damping wing absorption depends
only on the separation between the galaxy and the edge of the HII
region.\footnote{We assume pure Hubble flow in our calculations.
Infall onto collapsed halos inside the HII region
\citep{barkana04-inf,santos04} should have only a small effect on the
damping wing.}  It is thus straightforward to transform equation
(\ref{eq:probmbub}) into the distribution of $\tau_{\rm damp}$.  The
only subtlety is that the galaxy need not sit exactly in the center of
its ionized bubble.  We will assume for simplicity that galaxies are
uniformly distributed throughout the HII region, with the only
restriction that each be a distance $R_{\rm min}$ from the edge.  We
set $R_{\rm min}$ equal to the radius of the galaxy's HII region if
it were completely isolated, because its own photons guarantee that the
galaxy sits at least this distance from the nearest neutral gas.  Note
that the extreme assumption that all galaxies sit precisely in the
center of the ionized bubble changes our results by $\la 25\%$ (much
less in most cases).  The true distribution will be somewhere in
between these, because high-redshift galaxies are highly biased.

\begin{figure}
\begin{center}
\resizebox{8cm}{!}{\includegraphics{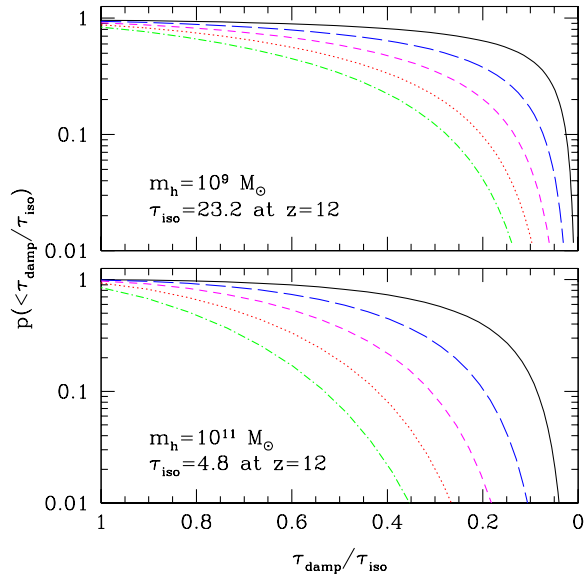}}\\%
\end{center}
\caption{The probability that halos with $m_h=10^9 \msun$ and
  $m_h=10^{11} \msun$ have a damping wing optical depth below some
  value in the $\zeta=40$ model.  In each case we parameterise
  $\tau_{\rm damp}$ in terms of the value it would have if only the
  galaxy's own HII region surrounded it, $\tau_{\rm iso}$.  The
  curves correspond to: $z=12$ ($\bxh=0.26$, solid), $z=13$
  ($\bxh=0.52$, long-dashed), $z=14$ ($\bxh=0.7$, short-dashed),
  $z=15$ ($\bxh=0.81$, dotted), and $z=16$ ($\bxh=0.89$, dot-dashed).}
\label{fig:ptaudamp}
\end{figure}

Figure \ref{fig:ptaudamp} shows the resulting distributions for the
same galaxies considered in Figure \ref{fig:raddamp}.  In each case we
scale $\tau_{\rm damp}$ to the value it would have if the galaxy were
completely isolated, $\tau_{\rm iso}$; the corresponding values at
$z=12$ are marked for the two halo masses.  (Note that $\tau_{\rm
damp}$ is $\sim 30\%$ higher at $z=16$; see equation
[\ref{eq:taudamp}].)  To compute $\tau_{\rm damp}$, we have assumed
that $\Delta z_d \equiv z_n-z_0=0.5$ and $\xh=1$ in this region.  The
true path length of neutral gas depends on the distance between
individual HII regions and will evolve as the bubbles grow and
overlap.  We have ignored this effect for simplicity because it does
not affect the main thrust of our argument.  We consider a more
detailed model in \S \ref{gap}.  Our choice here essentially maximises
the damping wing absorption because gas beyond $\Delta z_d=0.5$ is far
enough away in velocity space to render its absorption unimportant.
The simplest refinement would be to scale $\tau_{\rm damp}$ by the
mean neutral fraction, $\bxh$ (see \S \ref{abundance}).  Note that we
also neglect absorption by gas within the ionized bubble; we show
below that this resonant absorption is subdominant at the line center
(and especially on the red side) except when $\tau_{\rm damp} \la 1$.

The figure shows that there is a non-negligible probability of
substantially reduced absorption even early on, when $\bxh=0.9$: $\sim
20\%$ of $10^9 \msun$ galaxies have $\tau_{\rm damp}<15$ and $\sim
5\%$ have $\tau_{\rm damp}<6$.  As described above, the relative
effects are larger for smaller galaxies, because they are more likely
to have massive neighbors.  Of course, the absolute $\tau_{\rm damp}$
is still significantly smaller for larger galaxies.  If plotted as a
function of $\tau_{\rm damp}$, we note that the probability
distribution becomes less dependent on galaxy mass once the
characteristic bubble size exceeds $R_{\rm min}$.  

As emphasised above, the damping wing absorption depends on the IGM
characteristics.  If the IGM is partially ionized, either from an
earlier episode of reionization or from a smooth ionizing component
such as X-rays (e.g., \citealt{oh01b,venkatesan01}), $\tau_{\rm damp}$
decreases for two reasons.  Obviously, the intrinsic damping wing is
weaker if the external IGM is partially ionized.  This effect could be
accounted for by simply rescaling $\tau_{\rm iso}$ in Figure
\ref{fig:ptaudamp}.  However, the HII regions will also be larger for
a given ionizing luminosity, because fewer photons are consumed in
ionizing each IGM volume element.  As another example, we can consider
``double reionization'' scenarios.  An early generation of sources can
imprint a characteristic bubble size on the IGM that persists until
the second generation of sources has produced more ionizing photons
than the first generation \citep{furl04b}.  In this case the damping
wing absorption would be determined by the relic bubbles from the
first generation.  Thus the distribution of $\tau_{\rm damp}$ can
constrain complex reionization histories as well as the relatively
simple ones we have explicitly considered.

In principle, the damping wing absorption experienced by a galaxy
inside an HII region could be modified by its peculiar velocity.
However, this turns out to be unimportant: the typical velocity of a
halo at $z \sim 10$ is $\sim 100 \kms$ \citep{sheth01-vel}, much
smaller than the Hubble flow difference across its own Str{\" o}mgren
sphere.  Actually, the relative velocity shift between the galaxy and
the edge of the HII region is significantly smaller, because linear
theory peculiar velocities are determined by the mass distribution on
the largest scales and have fairly gentle gradients on Mpc scales.
Peculiar velocities could be more important for resonant absorption,
but nonlinear effects, including gas infall and internal motions such
as winds, are likely much more significant \citep{santos04}.

We see that the distribution of damping wing absorption is broad and,
especially during the middle and late stages of reionization, does not
directly map onto the ionizing output of an embedded galaxy.  Thus it
appears dangerous to attempt to constrain the ionization state of the
IGM outside of the Str{\" o}mgren sphere by comparing the inferred
absorption of the line to the estimated ionizing luminosity (even
neglecting the many complications intrinsic to the galaxy;
\citealt{santos04}).  Instead, the galaxy's neighbors play a crucial
role in determining the topology of ionized gas.  Thus we suggest that
attempts to infer the reionization history through observations of
individual high-redshift galaxies (e.g.,
\citealt{rhoads01,loeb04,cen04}) are highly uncertain.
For example, \citet{hue02} and \citet{rhoads04} claim that
observations of strong Ly$\alpha$ emission lines at $z \sim 6.5$
require the galaxies to be embedded in a mostly ionized medium.  Our
model predicts that a galaxy with $m_h = 10^{10} \msun$ has a
$(0.3\%,16\%,55\%)$ chance of $\tau_{\rm damp}<2$ if
$\bxh=(0.75,0.5,0.25)$ at $z=6.5$.  These become $(1\%,33\%,73\%)$
if $m_h=10^{11} \msun$.  Thus, it seems likely that $\bxh \la 0.5$,
but stronger conclusions are currently unwarranted.

On the other hand, our results suggest that searches for high-redshift
Ly$\alpha$ emitters can easily be extended to the era before
reionization.  While some fraction of galaxies will have their
Ly$\alpha$ lines completely absorbed by the surrounding IGM, many will
retain strong lines that can be isolated in narrow-band searches.
Most interestingly, the distribution of inferred damping wing optical
depths yields a measurement of the sizes of HII regions, which in
turn tells us about the morphology of reionization.  Of course,
we must account for effects intrinsic to the galaxy and its immediate
environments before strong constraints will emerge.

\begin{figure}
\begin{center}
\resizebox{8cm}{!}{\includegraphics{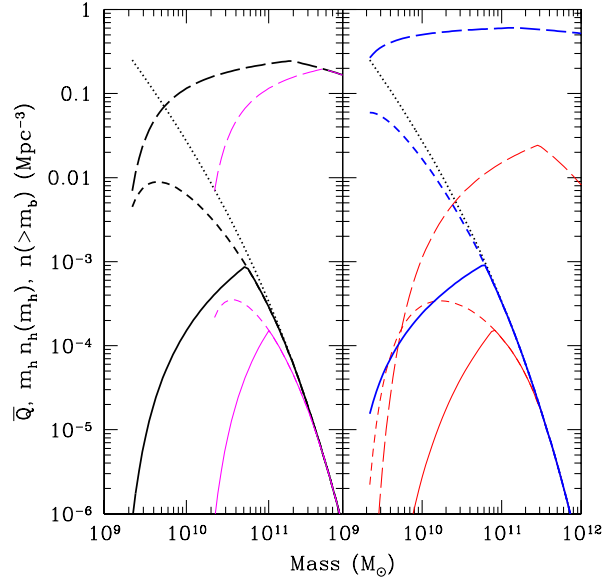}}\\%
\end{center}
\caption{ Cosmic variance in Ly$\alpha$ selected surveys at $z=8$.
  The dotted curves show the (logarithmic) halo mass function, $m_h
  n_h(m_h)$.  Short-dashed curves show the number density of halos
  that are visible given a minimum observable galaxy mass, $m_{\rm
  obs}$.  Solid curves show the approximate number density of HII
  regions of which these observable galaxies can be part, $n(>m_b)$,
  while the long-dashed curves show the fraction of space filled by
  these bubbles ($\Qb$).  The left panel assumes $\bxh=0.5$, with
  $m_{\rm obs}=10^{9}$ and $10^{10} \msun$ for the thick and thin
  curves, respectively.  The right panel assumes $m_{\rm obs}=10^{9}
  \msun$, with $\bxh=0.75$ and $0.25$ for the thick and thin curves.}
\label{fig:cosvar}
\end{figure}

One potential difficulty with surveys beyond reionization is that
cosmic variance could become extremely large, because the local
overdensity of galaxies modulates the damping wing absorption.  We
illustrate this point in Figure \ref{fig:cosvar}.  We first make the
simplest possible assumption that the Ly$\alpha$ line luminosity is
proportional to the galaxy mass, and we suppose that a given survey
can probe down to a minimum mass $m_{\rm obs}$, \emph{neglecting any
IGM absorption}.  This is obviously a poor assumption, because
galaxies of a single mass can have a wide range of star formation
rates, wind velocities, and intrinsic absorption, but it is probably
not too bad in the statistical sense.  We wish to find the fraction of
galaxies with mass $m_h$ that would be observable by such a survey and
how they are distributed in space.  Our assumptions require that a
galaxy have damping wing absorption $\tau_{\rm damp}$ smaller than
$\tau_{\rm max} \equiv \ln(m_h/m_{\rm obs})$ in order to be in the
survey. 

The dotted curves in Figure \ref{fig:cosvar} show the halo mass
function \citep{press}.  The short-dashed curves show the fraction of
galaxies at each mass that are accessible to our example survey, using
the formalism described above.  The solid curves show the number
density of HII regions that have $\tau_{\rm damp} < \tau_{\rm max}$
at their central points, while the long-dashed curves show the
fraction of space filled by such regions, $\Qb$.\footnote{Note that
these are only approximate calculations of the available number of
bubbles, because they require the galaxy to sit at the center of the
HII region (instead of any location within the bubble).}  The left
panel shows results for two different mass thresholds, $m_{\rm
obs}=10^{9}$ and $10^{10} \msun$ (thick and thin curves, respectively)
if $\bxh=0.5$ at $z=8$.  The number of visible sources is small near
the mass threshold, because the sources must be embedded in extremely
large bubbles, and they are confined to rare regions of the universe
(though the total number density is not small).  As $m_h$ increases,
however, the fraction of visible galaxies rapidly approaches unity and
the fraction of the universe through which they are distributed also
increases.  (Note that $\Qb$ decreases for sufficiently massive
galaxies because these galaxies are themselves so rare.)  This is
because the intrinsic luminosity is so large that they are visible
even when completely isolated.  The right panel shows results for
$m_{\rm obs}=10^9 \msun$ if $\bxh=0.75$ and 0.25 (thick and thin
curves, respectively).  Note the dramatic decline in the volume
available to the survey when $\bxh=0.25$; while some fraction of
galaxies are still visible, they are widely separated and highly
clustered.  Large-volume surveys will therefore be required,
especially in the earlier stages of reionization.

Finally, to further illustrate the importance of the size distribution of
bubbles, we now calculate some example emission line profiles.  We
will take as an illustrative example a galaxy with $m_h=10^{9} \msun$
at $z=10$.  Given a bubble size, we compute the damping wing
absorption from equation (\ref{eq:taudamp}).  However, in this case we
must also include the absorption by the (mostly ionized) gas near the
galaxy.  This gas is in ionization equilibrium, so (neglecting
clumping)
\bq
\alpha_B n_H^2(z) = \xh n_H(z) \bar{\sigma} (1+z)^2
\frac{\dot{N}_\gamma}{4 \pi R^2},  
\label{eq:ioneq}
\eq
where $\alpha_B$ is the case B recombination coefficient,
$\bar{\sigma} \approx 2 \times 10^{-18} \cmsq$ is the
frequency-averaged ionization cross-section, $R$ is the distance in
comoving units, and $\dot{N}_\gamma$ is the rate at which the central
source produces ionizing photons.  To compute this quantity, we
include only the ionizing photons from the galaxy (which dominate the
background at the relevant distances).  We assume a star formation
rate of $3 \sfr$, with the ionizing photon production rate given by
\citet{leitherer} for continuous star formation at $Z=0.05 Z_\odot$
and with a Salpeter initial mass function and a low-mass cutoff of
$M=1 \msun$.  We then allow a fraction $f_{\rm esc}=0.05$ of these to
escape the galaxy.  These parameters are of course arbitrary, but they
could describe the putative $z=10$ galaxy discovered by
\citet{pello04}.  For reference, these nominal parameters would
correspond to $\zeta=400 f_\star/(1+n_{\rm rec})$, where $f_\star$ is
the nominal star formation efficiency and $n_{\rm rec}$ is the mean
number of times a hydrogen atom recombines.  Finally, we assume that
the line width is $\sigma_v =V_c/\sqrt{2}$, where $V_c$ is the
circular velocity of the halo \citep{barkana01}; this yields
$\sigma_v=27 \kms$ for our parameters..

Equation (\ref{eq:ioneq}) yields the profile of $\xh$ near to the
galaxy.  To compute the resonant optical depth $\tau_{\rm res}$ at a
particular observed wavelength $\lambda_{\rm obs}$, we integrate the
Ly$\alpha$ absorption cross-section over this $\xh$ distribution,
assuming pure Hubble flow for the velocity structure.  We assume a
Voigt profile with $T=10^4 \kel$ for the absorption cross section, as
appropriate for ionized gas.  We note here that our treatment of
resonant absorption is highly simplified and neglects many important
effects.  For example, we have assumed the gas to be smoothly
distributed at the mean density; in reality it will have a complicated
density distribution.  Lines of sight passing through overdense
regions will have correspondingly more absorption; however, most of
space is filled by underdense gas and the mean resonant absorption is
somewhat weaker.  We estimate this effect in \S \ref{taueff} below.
We have also neglected infall and winds, which move the gas through
velocity space and can in principle strongly affect the absorption
\citep{santos04}.  Fortunately, neither of these effects change our
qualitative result, so we will not discuss them further here.

\begin{figure}
\begin{center}
\resizebox{8cm}{!}{\includegraphics{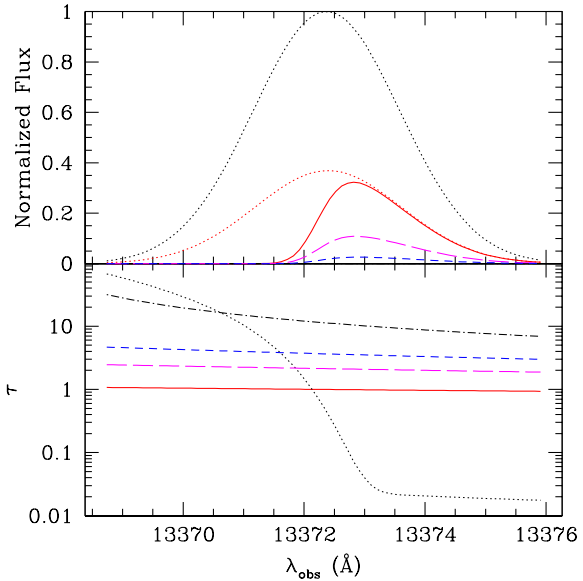}}\\%
\end{center}
\caption{\emph{Top panel:} Example line profiles for a galaxy at
  $z=10$.  The upper dotted curve shows the assumed intrinsic
  profile.  The solid, long-dashed, and short-dashed curves place the
  galaxy in HII regions of comoving size $R_b=10,\,5$, and $3 \Mpc$,
  respectively.  The lower dotted curve shows the absorption profile
  for $R_b=10 \Mpc$ if we neglect resonant absorption.  \emph{Bottom
  panel:} The solid, long-dashed, short-dashed, and dot-dashed curves
  show $\tau_{\rm damp}$ if the galaxy is in an HII region of
  comoving size $R_b=10,\,5,\,3$, and $1 \Mpc$, respectively.  The
  dotted line shows $\tau_{\rm res}$, which is the same for all bubble
  sizes. }
\label{fig:lp}
\end{figure}

Several example line profiles are shown in Figure \ref{fig:lp}, along
with the corresponding optical depths.  The solid, long-dashed,
short-dashed, and dot-dashed lines assume $R_b=10,\,5,\,3$, and $1
\Mpc$, respectively.  (The last of these is nearly completely absorbed
by the damping wing and does not appear in the top panel.)  In the top
panel, the upper dotted curve shows the assumed intrinsic line
profile; the other dotted line shows the absorption profile for
$R_b=10 \Mpc$ if we neglect resonant absorption.  In the bottom panel,
the dotted line shows $\tau_{\rm res}$.  For reference, an isolated
galaxy of this mass should have $R_b^{\rm iso} \approx
(\zeta/160)^{1/3} \Mpc$.  As a fiducial example, $\zeta=12$ implies
$\bxh=0.5$ at this redshift.  Then a galaxy with $m_h=10^9 \msun$
would have a $(75\%,\,40\%,\,20\%,\,3\%)$ probability of being in
bubbles larger than $(1,\,3,\,5,\,10) \Mpc$ according to our
formalism.  It is obvious that the large-scale environment of the
galaxy dramatically affects its absorption properties.  While the
resonant absorption will suppress the blue side of the line, it has
little effect on the red side, which allows an estimate of $\tau_{\rm
damp}$.

\section{Transmission Gaps in the Gunn-Peterson Trough}
\label{gap}

We now consider whether HII regions can appear as gaps in the
Gunn-Peterson troughs of distant quasars (unrelated to the ionized
bubbles that we wish to observe).  We will compute
\bq
\frac{dn(<\tau)}{dz} = \frac{dr}{dz} \int dm \, n(m) \pi R^2_{\rm
    max}(m,\tau),
\label{eq:dndz}
\eq 
where all distances are in comoving units, $dn(<\tau)/dz$ is the
number of transmission gaps with optical depths smaller than $\tau$
per unit redshift, and $n(m)$ is the mass function of ionized regions.
$R_{\rm max}$ is the impact parameter from the center of the HII
region at which the optical depth is $\tau$.  Note that the total
optical depth is twice the damping absorption (because both the blue
and red sides contribute) plus the resonant absorption.  We compute
the minimum optical depth encountered within each HII region (i.e.,
when the line of sight passes closest to the center of the bubble).

We first estimate the resonant absorption inside the ionized bubbles
in \S \ref{res}.  We then present the resulting absorption statistics
in \S \ref{abundance} and estimate how the inhomogeneous IGM affects
our results in \S \ref{taueff}.  Finally, we discuss the spectrum of
SDSS J1148+5251 in \S \ref{sdss}.

\subsection{Resonant Absorption}
\label{res}

Figure \ref{fig:lp} shows that resonant absorption is relatively
unimportant for Ly$\alpha$ emitters (at least on the red side of the
line).  This is because photons emitted on the red side never pass
through the Ly$\alpha$ resonance in the IGM and because the object we
are observing creates a local, highly-ionized bubble around itself.
The resonant absorption within this zone is suppressed because $\xh$
is much lower near the galaxy.  However, for lines of sight on a
\emph{random} path through an HII region, the probability of passing
so close to an ionizing source is small.  Thus most will suffer
substantial resonant absorption.

To compute $\tau_{\rm res}$, we therefore need to know the
distribution of ionizing sources within the HII region.  We will
begin by examining two simple limiting cases.  In the end, we will
show that they yield nearly identical results.  First, let us assume
that all of the ionizing sources are located in the center of the
bubble.  In that case, we can use equation (\ref{eq:ioneq}) once we
have an estimate for $\dot{N}_\gamma$.  Unfortunately, this factor
does not follow directly from the formalism described in \S
\ref{model}.  Our model uses the total time-integrated
number of ionizing photons (essentially our $\zeta$ parameter) to
compute the bubble sizes, while we need the instantaneous
photoionization rate to compute $\xh(R)$.  For simplicity, we will
assume that each HII region contains many sources, so that their
fluctuating star formation rates average out.  This is obviously only
accurate in the late stages of reionization, when the bubble sizes are
much larger than those of individual galaxies; fortunately, this
regime will turn out to be the relevant one anyway.  In this case, we
can set
\bq
\dot{N}_\gamma = \frac{ \zeta m_b}{\mu m_H} \dot{f}_{\rm coll}
\label{eq:ndotgamma}
\eq
for a bubble of mass $m_b$.  Then
\bqa
\xh^c(R) & = & \frac{3 \alpha_B}{\bar{\sigma} R_b H(z)} \ \left(
\frac{R}{R_b} \right)^2 \ \left( \frac{f_{\rm coll}}{|df_{\rm
    coll}/dz|}  \right) \label{eq:xHIcenter} \\
& \approx & 3.2 \times 10^{-3} \, \left(  \frac{{\rm Mpc}}{R_b} \right) \
\left( \frac{R}{R_b} \right)^2 \ \left( \frac{10}{1+z} \right)^2
\nonumber \\
& & \times \ \left( \frac{f_{\rm coll}}{|df_{\rm coll}/dz|}  \right) \ \left(
\frac{0.15}{\Omega_m h^2} \right)^{1/2}, \nonumber
\eqa
where the ``c'' superscript refers to the case in which all the
ionizing sources are at the center of the bubble.  Note that the
$f_{\rm coll}$ factor is typically of order 3 at these redshifts.  The
resonant absorption at the specified velocity is then $\tau_{\rm res}
\approx \tau_{\rm GP} \xh$ and will be quite large except near the
center of the regions.  We have ignored one subtlety in equation
(\ref{eq:ndotgamma}): in general $\zeta$ includes the mean number of
recombinations $n_{\rm rec}$ in the IGM, integrated over the entire
reionization history to this point.  The instantaneous ionizing rate,
of course, does not depend on the recombination rate in the past. Thus
$\xh^c \propto (1+n_{\rm rec})^{-1}$; in a scenario in which
recombinations have delayed reionization but the mean ionizing
emissivity is high, this could have a substantial effect.

As a second limiting case, we assume that the sources are distributed
uniformly throughout the bubble.  In this case the flux of ionizing
photons is also approximately uniform across the HII region
(non-uniformity occurs only because the bubble has a finite size); in
equation (\ref{eq:ioneq}), we have
\bq
\frac{\dot{N}_\gamma}{4 \pi R^2} \rightarrow \frac{3 \dot{N}_\gamma}{4
  \pi R_b^2}. 
\label{eq:fluxu}
\eq
The resulting ionized fraction is
\bq
\xh^u = \xh^c(R_b)/3.
\label{eq:xHIu}
\eq
We find that the resonant absorption is $\tau_{\rm res} \sim 10^{-3}
\tau_{\rm GP}(z) ({\rm Mpc}/R_b)$ at a typical position within a bubble,
or $\tau_{\rm res} \ga 60$ for $R_b=10 \Mpc$.  Thus the mean neutral
fraction is large, so $\tau_{\rm res} \gg 1$ unless the line of sight
passes close to an ionizing source.  This has a fortunate consequence:
the cross-section for a line of sight with $\tau_{\rm res}<\tau_0$ is
\emph{independent} of the number of sources (see also
\citealt{barkana02-bub}).  Suppose that the bubble is made up of $N$
identical galaxies.  From equation (\ref{eq:ioneq}), the radius at
which the optical depth equals a fixed value $\tau_0$ is $R_{\rm
max}(\tau_0) \propto 1/\sqrt{N}$.  Thus the total cross-section
through such regions is $N \times \pi R_{\rm max}^2 =$ constant and is
independent of the number of ``subclusters'' within the HII region.
A similar argument shows that the cross-section also remains unchanged
if only a fraction of the bubbles are ``active'' at a given time.  Of
course, if the galaxies or subclusters are distributed throughout the
bubble, their damping wing absorption will change with distance from
the edge.  Because this is a relatively small effect, we will use
equation (\ref{eq:xHIcenter}) to compute the resonant absorption
inside the HII regions.

Note that the above treatment assumes a uniform density IGM.  We will
show that clumping actually \emph{increases} $R_{\rm max}$ in \S
\ref{taueff}.

Both of the above prescriptions neglect absorption of the ionizing
photons as they travel through the bubble.  This requires the total
optical depth to ionizing photons $\tau_{\rm ion}$ to be $\la 1$.  We
find that 
\bqa
\tau_{\rm ion}(R_b) & = & \frac{n_H(z) \alpha_B}{(1+z) H(z)} \ \left(
\frac{f_{\rm coll}}{|df_{\rm coll}/dz|} \right) \\
& \approx & 0.1 \left( \frac{f_{\rm coll}}{|df_{\rm coll}/dz|}
\right) \ \left( \frac{1+z}{10} \right)^{1/2} \nonumber \\ 
& & \times \ \left(
\frac{0.15}{\Omega_m h^2} \right)^{1/2} \left( \frac{\Omega_b
  h^2}{0.023} \right), \nonumber
\label{eq:taubub}
\eqa
in both the centrally concentrated and uniform limits.  Thus,
neglecting any dense clumps, the majority of ionizing photons reach
the edge of the bubble without being consumed by recombinations.  This
is, in fact, a necessary assumption of the model described in \S
\ref{model}.  Of course, if a large fraction of lines of sight pass
through dense ``Lyman-limit'' systems before hitting the edge of the
bubble (or in other words if the characteristic bubble size exceeds
the mean free path between such systems), the HII region would no
longer expand and our model breaks down.  The point at which this
occurs is of course unknown; note that it could be quite early if
``minihalos'' are common \citep{barkana02}.

\subsection{The Abundance of Transmission Gaps}
\label{abundance}

We are now in a position to compute $dn(<\tau)/dz$ from equation
(\ref{eq:dndz}).  We calculate the resonant absorption via the
centrally concentrated limit of equation (\ref{eq:xHIcenter}) and
$\tau_{\rm res} \approx \xh \tau_{\rm GP}$.\footnote{Note that we
neglect the spatial variation of $\xh$ here.  In principle, we should
compute the mean value of $\xh$ smoothed across a thermal width of the
line, but the width is usually small compared to the length scales of
interest.}  For the damping wing, we must know the total path length
of the neutral segment of gas (i.e., $z_n - z_0$ in equation
[\ref{eq:taudamp}]).  This is difficult to compute precisely, because
it depends on the large-scale distribution of the HII regions.  We
will take $\Delta z_d=0.5$ and assume that the neutral fraction is
$\bxh$ within this region.  We examine the importance of this
prescription below (where we also show some examples of $R_{\rm
max}$).

\begin{figure}
\begin{center}
\resizebox{8cm}{!}{\includegraphics{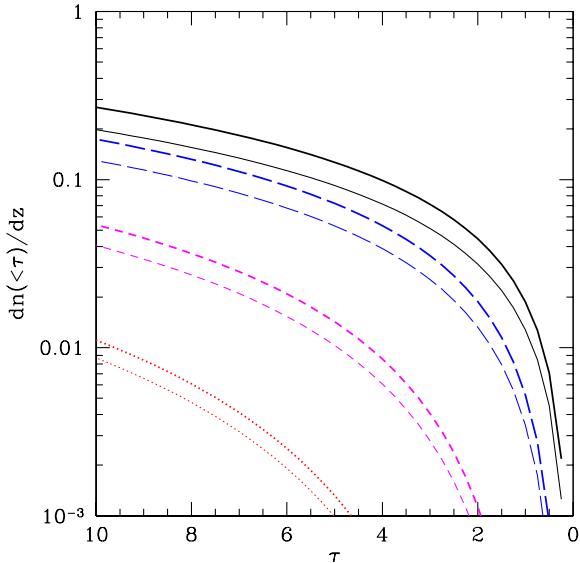}}\\%
\end{center}
\caption{ The number of transmission gaps per unit redshift $dn/dz$
  with an optical depth smaller than the specified value.  The thick
  curves assume $\zeta=12$ and have $\bxh=0.12$ ($z=8.5$, solid),
  $\bxh=0.3$ ($z=9$, long-dashed), and $\bxh=0.5$ ($z=10$,
  short-dashed), $\bxh=0.66$ ($z=11$, dotted).  The thin curves assume
  $\zeta=40$ at the same neutral fractions; these occur $\Delta z \sim
  3$ earlier than the thick curves. }
\label{fig:lya}
\end{figure}

Figure \ref{fig:lya} shows how the absorption statistics evolve
through reionization in our model.  The thick and thin curves assume
$\zeta=12$ and $40$, respectively.  The differences between the two
are relatively modest, despite being displaced in redshift by $\Delta
z \sim 3$, and occur because the higher cosmic densities in the
$\zeta=40$ model make absorption somewhat stronger.  The curves are
fairly flat until some characteristic optical depth is reached; this
is because the bubbles themselves have a characteristic size.  The
curves flatten with time because resonant absorption becomes less
significant as the bubbles grow and because the peak in the bubble
size distribution sharpens with time.  The expected number of
absorbers is clearly quite small; even when $\bxh \sim 0.1$, only
$\sim 0.1$ gaps with $\tau \la 4$ are expected per unit redshift.
However, in contrast to \citet{barkana02-bub}, the probability is not
vanishingly small unless $\bxh \ga 0.5$.  Even a few quasars shortly
beyond reionization could show some transmission features.  The
difference arises because \citet{barkana02-bub} considered isolated
HII regions around galaxies, for which the damping wing absorption
provides a relatively large ``floor'' in the optical depth.  In our
model, the damping wing absorption can be much smaller at the centers
of large bubbles.  Finally, we have set $n_{\rm rec}=0$ when computing
the resonant absorption (see the discussion of equation
[\ref{eq:xHIcenter}]).  If recombinations have significantly delayed
reionization, $dn/dz$ will increase approximately in proportion to
$1/(1+n_{\rm rec})$, except at the bright end where the damping wing
is important.

We note here that it may be easier to detect regions of weak
absorption in the Ly$\beta$ Gunn-Peterson trough, because $\tau_\beta
= 0.16 \tau_\alpha$; the optical depth axis in Figure \ref{fig:lya} can
simply be scaled by this amount.  We therefore expect a factor of
several more such gaps near the end of reionization and orders of
magnitude more during the early stages.  Of course, the Ly$\beta$
region is blanketed by the lower-redshift Ly$\alpha$ forest, which
will cover some fraction of the transmission gaps (especially since
the Ly$\alpha$ forest should still be fairly deep at the relevant
redshifts).

Assuming a uniform density IGM, we can easily estimate the widths of
the transmission gaps.  In the regime in which resonant absorption
dominates, the optical depth will increase by a factor of two over a
distance along the line of sight equal to the impact parameter:
$\Delta \lambda/\lambda_{\rm obs} \sim H(z) R_{\rm max}/c \sim 0.003
[(1+z)/10]^{3/2} (R_{\rm max}/{\rm Mpc})$.  Interestingly, the line
widths can thus break the degeneracy between a small number of
clustered sources and a large number of distributed sources.  As
described above, the total cross section does not change if we
separate the central ionizing source into $N$ small clusters, because
$R_{\rm max} \propto N^{-1/2}$.  However, each individual line of
sight must pass closer to a source if clustering is weak, so the line
widths would decrease.  Similarly, if the central sources of HII
regions are bursty, each ``active'' region would have a larger $R_{\rm
  max}$, making the individual lines larger.  Note, however, that the
optical depth gradients are much smaller if $\tau_{\rm damp}$ is
important. 

\begin{figure}
\begin{center}
\resizebox{8cm}{!}{\includegraphics{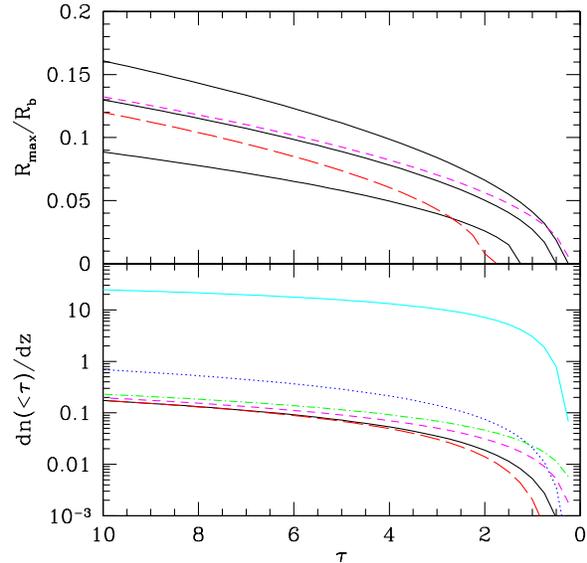}}\\%
\end{center}
\caption{ \emph{Top panel:} The cross-section for bubbles.  The solid
  curves have $R_b=15,\,10$, and $5 \Mpc$, from top to bottom.  The
  long and short dashed lines modify the damping wing (see text) and
  have $R_b=10 \Mpc$.  \emph{Bottom panel:} Density of transmission
  gaps.  The lower solid curve is our default model, and the long and
  short dashed lines modify the damping wing.  The dotted line assumes
  that the resonant absorber has $\Delta = 0.5$.  The upper solid line
  includes only absorption from the damping wing, while the dot-dashed
  line includes only resonant absorption.  All curves (in both panels)
  assume $\zeta=12$ and are at $\bxh=0.3$ ($z=9$).  }
\label{fig:lyaparams}
\end{figure}

Figure \ref{fig:lyaparams} shows how our results depend on some of our
input parameters.  The top panel shows $R_{\rm max}$ if $\zeta=12$ and
$z=9$ (at which time $\bxh=0.3$).  The three solid curves correspond
to $R_b=15,\,10$, and $5 \Mpc$, respectively.  In all cases, the
impact parameter must be a small fraction of the total bubble size in
order for the local ionizing background to be large enough to reduce
the resonant absorption.  The ratio $R_{\rm max}/R_b$ increases with
$R_b$ both because the damping wing is farther away and because the
source density increases.  These curves change weakly with redshift
because of the increased absorption in a denser universe.  The other
curves show how the results change with different assumptions about
the damping wing.  To construct the long-dashed curve, we assumed a
\emph{fully} neutral medium with $\Delta z_d=0.5$.  The larger
$\tau_{\rm damp}$ induces a more significant cutoff at the bright end,
but the effects are not large at moderate optical depths because
resonant absorption dominates.  The short-dashed curve uses a
prescription similar to that of \citet{barkana02-bub}.  We let
$dn_{\rm ion}/dz$ be the number of ionized bubbles intersected per
unit redshift; this is the same as equation (\ref{eq:dndz}) except
that $R_{\rm max} \rightarrow R_b$.  The mean path length of a neutral
segment (in redshift units) is then given by $\bxh/\Delta z_d \sim
{dn_{\rm ion}/dz}$.  The damping wing absorption declines in this
model because the path length between ionized bubbles is small.
However, when $\bxh$ is small the prescription makes only a small
difference because the damping wing is unimportant anyway.

The bottom panel of Figure \ref{fig:lyaparams} shows the effects on
the transmission gap statistics.  The lower solid curve is the normal
model of Figure \ref{fig:lya}, while the dashed curves make the same
assumptions as in the top panel.  The effects on the gaps are clearly
modest, except at the bright end.  However, note that our assumptions
about $\tau_{\rm damp}$ can be much more important for $\bxh \ga 0.5$,
because even at that time $dn/dz_{\rm ion}$ can be relatively large
(although the path length through each ionized bubble is small).
Nevertheless, the number of observable transmission gaps is extremely
small in this regime, even in the most optimistic scenario.

The upper solid curve is identical to our fiducial model except that
we include \emph{only} the damping wing component in the absorption.
The amplitude is much larger in this case: as argued above, resonant
absorption is the principal limiting factor in finding gaps.
Comparing the small $\tau$ tails of the distributions, we see that the
cutoff in the number of absorbers is fixed by the ``floor'' in damping
wing absorption.  In contrast, the dot-dashed curve includes only
resonant absorption.  The number density is not far from the full
model (except at the bright end), indicating that this quantity
controls the gap abundance.  This has important implications for
interpreting quasar spectra in light of complex reionization histories.
Figure \ref{fig:lya} assumes a simple model with a single generation
of sources.  The dominance of resonant absorption implies that the
statistics will be qualitatively unchanged for more complex models,
although we would expect a substantial increase in the number of
nearly transparent gaps (with $\tau<1$).  For example, if the external
IGM is partially ionized, $\tau_{\rm damp}$ will decrease and the
bubbles will be larger for a given ionizing emissivity.  However, the
resonant absorption within the bubbles will remain the same, and the
dot-dashed curve shows an \emph{upper} limit on the resulting number
of gaps.  Similarly, although an early generation of sources can
increase the bubble size, it does not affect the resonant absorption.
Thus we expect that $dn/dz$ places stronger constraints on the
instantaneous ionizing emissivity than on the total ionized fraction,
as long as the HII regions are large enough to reduce $\tau_{\rm
damp}$ to a subdominant role.  This could be especially interesting
when combined with other methods (such as Ly$\alpha$ emitters or 21 cm
tomography) that are more sensitive to the total ionization history.

\subsection{The Effect of an Inhomogeneous IGM}
\label{taueff}

The most important simplification we have made is to neglect the
density distribution of the IGM.  On the one hand, small-scale
clumpiness increases the mean recombination rate and hence the
resonant absorption inside bubbles.  On the other hand, underdense
regions with smaller optical depth fill most of the volume, so the
majority of any line of sight is likely to pass through regions with
smaller absorption.  As a quick illustration, the long-dashed curve in
Figure \ref{fig:lyaparams} shows how the results change if we assume
that the gas causing resonant absorption has $\Delta=0.5$, where
$\rho=\Delta \bar{\rho}$.  Equation (\ref{eq:ioneq}) shows that in
ionization equilibrium the local neutral fraction is proportional to
the local density, and the optical depth has an extra power of
density.  Thus $\tau_{\rm res} \propto \Delta^{2}$ at a fixed distance
from the center of the bubble. This is close to the behavior we find
in Figure \ref{fig:lyaparams} at moderate $\tau$, though the
amplification is smaller at $\tau<1$ because the damping wing becomes
important.

This is only the crudest estimate of the importance of density
fluctuations.  To do better, we must average the magnitude of
absorption over the density structure of the IGM: we wish to compute
$\tau_{\rm eff} \equiv - \ln \langle e^{-\tau} \rangle$ rather than
simply $\langle \tau \rangle$ as we did above.  The best way to
approach the problem is with numerical simulations, but these are
computationally expensive and cannot yet be performed with the
necessary dynamic range.  Here we take an approximate approach.  We
wish to estimate the fluctuations in the density field when smoothed
along the line of sight over a thermal width of the absorbing gas.
\citet{miralda00} showed that the volume-averaged probability
distribution of gas overdensity in numerical simulations at $z \sim
2$--$4$ is well fit by
\bq
P_V(\Delta) d \Delta = A \Delta^{-\beta} \exp \left[
  \frac{-(\Delta^{-2/3} - C)^2}{2 (2 \delta_0/3)^2} \right] d \Delta
\label{eq:pdelta}
\eq
when smoothed on the Jeans mass of the photoionized gas.  The form is
motivated by interpolating between linear theory and the growth of
voids and halos in the nonlinear regime.  The parameter $\delta_0$
describes the linear regime: \citet{miralda00} find that it is
given by $\delta_0=7.61/(1+z)$.  The exponent $\beta$ depends on
the density profile of collapsed objects; we will set $\beta=2.5$, the
value it takes for isothermal spheres.  We then fix the remaining
parameters $A$ and $C$ by normalizing the mass and volume-weighted
probability densities.  Given the density distribution, we can compute
the effective volume-averaged optical depth, $\tau_{\rm eff}$,
corresponding to a nominal optical depth for the mean density IGM, 
$\tau_{\rm mean}$, through 
\bq 
\langle e^{-\tau} \rangle = \int d\Delta \, P_V(\Delta) \exp
(-\tau_{\rm mean} \Delta^2 ). 
\label{eq:taueff}
\eq

Unfortunately, there are several reasons why this treatment is only
approximate.  First, \citet{miralda00} used simulations with
$\Omega_m=0.4$, $\Omega_\Lambda=0.6$, and $\sigma_8=0.79$, so there
should be minor differences with the cosmology we have chosen.
Moreover, the simulation did not resolve fluctuations on small
scales: while the Jeans mass at $T=10^4 \kel$ is resolved,
fluctuations on smaller scales are not.  Such small scale clumpiness
is probably not important at $z \sim 3$, for which the simulations
were designed, because the thermal pressure of the ionized gas would
have smoothed any density fluctuations.  However, it may underestimate
the fluctuations in recently ionized gas that has not yet relaxed into
pressure equilibrium.  Minihalos are an important example that could
drastically increase the mean recombination rate until they
photoevaporate \citep{barkana02,shapiro04}.  Furthermore, equation
(\ref{eq:pdelta}) smooths over the Jeans scale, which is slightly
smaller than the thermal broadening width relevant for absorption
studies.  (However, we should also perform the average over one
dimension rather than three.)  Finally, equation (\ref{eq:pdelta})
does not take into account the environment of the volume element under
consideration.  For resonant absorption, we are primarily concerned
with regions that are near galaxies or groups of galaxies; we would
therefore expect that the relevant regions would be less likely to
contain a void.  It is not clear which of these effects dominate, so
we simply take equation (\ref{eq:pdelta}) at face value and caution
the reader that our results are only approximate.

\begin{figure}
\begin{center}
\resizebox{8cm}{!}{\includegraphics{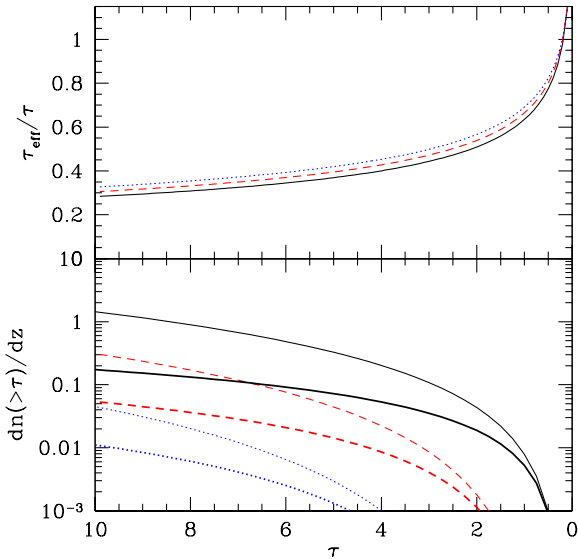}}\\%
\end{center}
\caption{ The top panel shows the ratio of $\tau_{\rm eff} = - \ln
  \langle e^{-\tau} \rangle$ to the optical depth of the mean density
  IGM, $\tau$, at $z=9,\,10$, and $11$ (solid, dashed, and dotted
  curves, respectively). The bottom panel shows the number of
  transmission gaps per unit redshift $dn/dz$.  The thick curves
  assume that all gas has the mean density, while the thin curves
  account for the density distribution of the gas as in the top panel.
  All the curves assume $\zeta=12$, which has $\bxh=0.3$ ($z=9$),
  $\bxh=0.5$ ($z=10$), and $\bxh=0.66$ ($z=11$).  }
\label{fig:taueff}
\end{figure}

We show the results of equation (\ref{eq:pdelta}) in the top panel of
Figure \ref{fig:taueff} for three different redshifts, $z=9,\,10$, and
$11$ (from bottom to top).  We see that $\tau_{\rm eff}$ decreases
slowly as redshift decreases because structures condense and voids
fill up more of the volume.  The decrease in apparent optical depth is
a factor of $\sim 3$--$4$ at moderate $\tau$.  The difference becomes
negligible for small $\tau$ (and close enough to zero, we even have
$\tau_{\rm eff}>\tau_{\rm mean}$) as the extra clumpiness becomes
important.  The correction to $\tau$ is significantly smaller than
that of \citet{haiman02}, who assumed a much clumpier density
distribution.

The bottom panel of Figure \ref{fig:taueff} compares the $dn/dz$
distributions for a uniform IGM (thick lines) with those using
$\tau_{\rm eff}$ (thin lines).  As expected from Figure
\ref{fig:lyaparams}, the differences are substantial (up to an order
of magnitude at $\tau \sim 10$), although they become much less
important at small optical depths.  On the other hand, the number of
transmission gaps remains small but non-negligible, so our fundamental
conclusions are not affected.  Additionally, we emphasise that our
method may miss extra small-scale clumpiness and the environmental
bias.  We conclude that transmission gaps are likely to be rare before
reionization, but their appearance near the end of overlap is by no
means impossible.

We note that the density distribution does not have a strong effect on
the damping wing, because equation (\ref{eq:taudamp}) averages over a
large path length through the IGM.  Also, the $\tau_{\rm eff}$ curves
in the bottom panel of Figure \ref{fig:taueff} cannot be scaled to
$\tau_\beta$, because the correction in equation (\ref{eq:taueff})
must then be evaluated with $\tau_{\beta,\,{\rm mean}}$ in the
exponent.

\subsection{The Transmission Gap in SDSS J1148+5251}
\label{sdss}

To date, SDSS has found four quasars with apparent Gunn-Peterson
troughs, all at $z>6.2$ \citep{fan,becker01,fan03,white03,fan04}.
Together, the four troughs have a total path length of $\Delta z \sim
1$.  SDSS J1148+5251 has a transmission spike at 8609 \AA ($z=6.08$
for a Ly$\alpha$ line) with an apparent $\tau \approx 2.5$, as well as
a matching Ly$\beta$ peak \citep{white03}.  White et al. point out
that such a gap is not hard to imagine if one includes only the
damping wing; however, they neglected resonant absorption.  Because it
treats each galaxy in isolation, the model of \citet{barkana02-bub}
essentially requires that this gap appears only well after overlap.

\begin{figure}
\begin{center}
\resizebox{8cm}{!}{\includegraphics{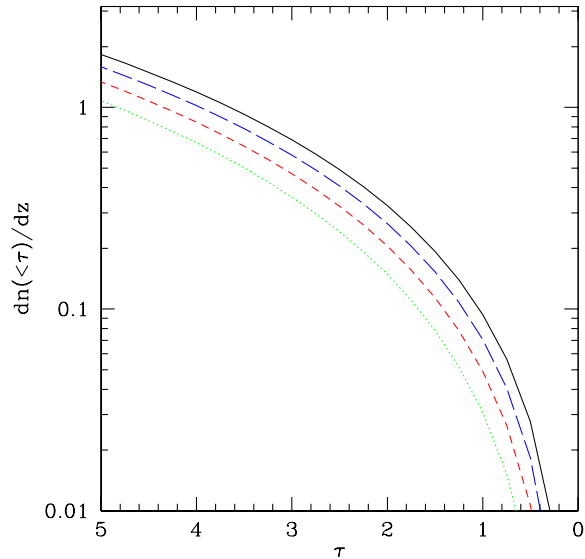}}\\%
\end{center}
\caption{ The expected number of transmission gaps at $z=6.1$ if the
  universe has $\bxh=0.1,\,0.15,\,0.2$, and $0.25$ (solid,
  long-dashed, short-dashed, and dotted curves, respectively).  We
  have used the inhomogeneous IGM model described in \S \ref{taueff} to
  construct the curves. }
\label{fig:sdss}
\end{figure}

As we have emphasised, our large HII regions reduce the severity of
the damping wing and allow transmission spikes near the end of
overlap.  Figure \ref{fig:sdss} shows the expected transmission
statistics for $\bxh=0.1,\,0.15,\,0.2$, and $0.25$ (top to bottom) at
$z=6.1$.  We have used our standard assumptions plus the inhomogeneous
IGM model of \S \ref{taueff} to construct the curves.  We see that
$dn(\tau<2.5)/dz \sim 0.25-0.5$ for these neutral fractions.  This is
entirely consistent with the observed gap; thus we argue that its
existence does not \emph{require} that overlap occurred before
$z=6.1$.  Of course, it is still perfectly consistent with such a
scenario.

We caution the reader against overinterpreting these results.
Uncertainties about the true distribution of sources and the density
structure of the IGM make a quantitative comparison difficult at
present (especially with the small observational sample).  Moreover, our
model probably begins to break down near overlap:  by this point, the
bubbles are so large that small scale structure and Lyman-limit
systems will play an important role in determining the ionizing
background.   

Finally, the same quasar spectrum shows other Ly$\beta$ transmission
gaps that could be compared to our model.  However, the line of sight
also contains a CIV absorber at $z=4.93$, and some of these
spikes are likely emission lines from the same system \citep{white03}.
More data about which gaps are truly due to the interloper are crucial
for extracting the most information possible from this spectrum.

\section{Discussion}
\label{disc}

We have considered two methods to study the topology of reionization
through near-infrared observations of high-redshift sources.  The most
promising technique is to measure the effects of damping wing
absorption on the Ly$\alpha$ emission lines of high-redshift galaxies.
Narrowband searches for galaxies with strong lines are one useful
technique, and these surveys currently extend to $z \sim 6.5$ (e.g.,
\citealt{hue02,kodaira03,rhoads04,stanway04,santos04-obs}).  They become
increasingly difficult at higher redshifts because of the sky
background, but some narrow transmission windows do remain in the
near-infrared, and we expect high-redshift galaxies to have strong
intrinsic Ly$\alpha$ lines (e.g., \citealt{partridge67,barton04}).  Of
course, the technique need not rely on galaxies selected in this way
and can be applied to broadband photometric dropouts (e.g.,
\citealt{stanway04-goods}) as well as more specialised selection
techniques (e.g., gravitational lensing, \citealt{kneib04,pello04}).

Our results have several implications for such surveys.  First,
Ly$\alpha$ selection should be able to extend to eras well before
reionization.  Although at these times many galaxies will suffer
significant damping wing absorption, reducing the apparent strength of
their Ly$\alpha$ emission lines, a substantial fraction of galaxies
will sit inside of large HII regions that allow transmission of a
significant amount of flux in the Ly$\alpha$ line even when $\bxh \sim
0.75$.  We thus expect that the number density of narrowband-selected
galaxies will decline with redshift but that they will not disappear
entirely.  Second, because of the wide variance in IGM absorption for
galaxies with the same intrinsic properties, we find that a single
galaxy cannot provide tight constraints on either the ionization state
of the IGM or the properties of the host galaxy
\citep{hue02,rhoads04,loeb04,ricotti04,cen04}.  Including the galaxy's
neighbors can make a dramatic difference to the expected level of
inferred absorption.  For example, massive galaxies at $z=6.5$ have a
$\ga 20\%$ probability of sitting in HII regions with $\tau_{\rm damp}
< 2$ even if $\bxh=0.5$.  Fortunately, the most massive galaxies
generally dominate their local HII regions, so in that case such
inferences will not be as dangerous (see Figure \ref{fig:ptaudamp}).
Finally, our model also shows that cosmic variance in
Ly$\alpha$-selected surveys will be even greater than expected,
because the local overdensity of galaxies modulates the damping wing
absorption (see Figure \ref{fig:cosvar}).  We predict that, although
some galaxies remain visible if $\bxh \la 0.5$, they are confined to
increasingly rare regions of the universe.

On the other hand, the wide range of damping wing absorption expected
through the middle ranges of reionization allows us to constrain the
size distribution of HII regions during this era.  This in turn will
teach us about the properties of both the IGM and the high-redshift
sources.  Interestingly, because partial uniform reionization (such as
from X-rays; \citealt{oh01b,venkatesan01}) and double reionization
(e.g., \citealt{wyithe03,cen03,haiman03}) affect the bubble sizes,
such surveys can also help to constrain these models.  Ideally, such a
survey would use galaxies that are \emph{not} selected through
narrow-band imaging to avoid the selection effect described above, but
for the time being Ly$\alpha$ line surveys look to be the most
promising technique (at least for large cosmological volumes).  Of
course, we must separate the damping wing absorption from a variety of
other complications, including gas infall, small-scale clumpiness, and
galactic winds \citep{santos04} before strong constraints can be made.
With large samples, this should not be too difficult because we do not
expect these local features to differ substantially across the
large-scale environments that determine the topology of ionized
bubbles.

A related technique is to measure the damping wing absorption from a
gamma-ray burst afterglow \citep{miresc98,barkana04-grb}.  The
advantage of these objects is that their smooth power-law spectra (and
lack of Ly$\alpha$ emission lines) makes interpreting the damping wing
much simpler.  Follow-up spectroscopy could then measure the host
galaxy's redshift and comparison to the damping wing would yield the
size of the HII region.  These host galaxies would be selected in an
unbiased manner relative to the Ly$\alpha$ absorption, which would
make their interpretation easier.

Of course, our model is only approximate.  Numerical simulations are
required to address questions such as the anisotropy of HII regions,
the resonant absorption, and the effects of peculiar velocities.
\citet{gnedin04} have already made some steps in this direction;
however, our model suggests that their box (with size $8 h^{-1}$
comoving Mpc) is probably too small to obtain precise measurements,
and in any case they were only able to examine one specific
reionization history.

A second method to study the topology of reionization is to examine
high-resolution quasar absorption spectra, which could contain gaps in
the Gunn-Peterson trough when the line of sight passes close to the
center of a large HII region.  Previous studies had suggested that the
strong damping wing absorption by the surrounding neutral IGM makes
such gaps extremely rare.  We have shown that the clustering of
high-redshift galaxies makes the ionized bubbles much larger than
naively expected, reducing the contribution of the damping wing.  In
this case, gaps are still expected to be rare because residual neutral
gas (even with $\xh \sim 10^{-5}$) can provide substantial resonant
absorption; late in reionization we expect $dn/dz \sim 0.02$--$0.3$
for features with $\tau < 2$.  However, weak transmission gaps are not
so rare that their presence automatically implies an earlier episode
of reionization.  In particular, the gap at $z=6.08$ in the spectrum
of SDSS J1148+5251 \citep{white03} does not necessarily imply an
earlier overlap epoch (though of course it is consistent with such a
scenario).  It is encouraging that existing and near-future
observations can help to constrain the topology of reionization.
Because this technique relies on active ionizing sources to eliminate
the resonant absorption, it is not terribly sensitive to complex
reionization histories (which mostly affect the damping wing).  Large
HII regions already reduce the damping wing to manageable levels.
Instead quasar absorption spectra are most sensitive to the
instantaneous ionizing rate.  We have also shown that the number of
transmission gaps is nearly independent of the clustering and
burstiness of ionizing sources.  However, the widths of the gaps do
vary with these quantities, so measurements of their sizes will help
to constrain these parameters.

The principal source of uncertainty in this calculation is the
inhomogeneous density distribution of the IGM, which we have accounted
for only approximately.  Unfortunately, the clumpiness of the IGM is
difficult to quantify analytically.  Numerical simulations can do much
better, but we note that resolving the small scale density structure
is particularly difficult in this case because the transmission gaps
occur in the rare, large HII regions with sizes $\ga 10 \Mpc$
(comoving).  A complete solution would therefore require a large box
and high mass resolution.  Moreover, our model for the HII
regions is only approximate and breaks down somewhat before overlap,
once the mean free path of a photon is determined not by the bubble
size but by small-scale overdensities (i.e., Lyman-limit systems).
The topology of reionization during this era is best addressed through
numerical simulations.

We also note that transmission gaps can occur if the line of sight
passes near to a bright quasar \citep{miresc98,miralda00}.  Quasars
produce such strong ionizing radiation fields ($\dot{N}_\gamma \ga
10^{57}$ photons s$^{-1}$ for the bright SDSS quasars, assuming the
\citealt{elvis94} template spectrum) that $\tau_{\rm res}$ is small
within several comoving Mpc of the quasar.  Estimates of the number
density of gaps are difficult because they depend on the unknown
luminosity function of fainter quasars.  The best current estimate for
quasars with $M_{1450}<26.7$ is $n_{\rm QSO} \sim 6 \times 10^{-10}
\Mpcden$ at $z \sim 6$ \citep{fan04}.  For these quasars, we then have
$dn/dz \sim 10^{-4}$ for $\tau \la 1$ even if we neglect damping wing
absorption.  Thus the currently observed population should provide a
negligible number of intersections.  Extrapolating to fainter quasars
is dangerous given the lack of observational constraints; however, if
the luminosity function has a power law slope $\beta \sim -3$
\citep{fan04} at the bright end, this steep component would have to
extend to quasars a factor of $>100$ fainter than the observed
population in order for gaps due to quasars to become comparable to
our predictions for $\bxh < 0.3$.  This would require the break
luminosity to occur at significantly smaller luminosities than
observed at $z\sim2$ \citep{boyle00}.  Note also that the number
density of quasars declines rapidly with redshift (it drops by a
factor $\sim 30$ from $z=3$ to $z=6$; \citealt{fan04}), so quasars
likely become even less significant at higher redshifts.
Finally, because quasar HII regions are so large, the gradients in
resonant absorption are relatively gentle and gaps near quasars should
be quite wide in velocity space.

In summary, we find that both detailed studies of the absorption
suffered by the Ly$\alpha$ emission lines of star-forming galaxies and
searches for transmission gaps in the spectra of high-redshift quasars
will enable us to constrain the size distribution of HII regions in
the middle and late stages of reionization.  Both of these methods are
already able to probe the tail end of reionization at $z \sim 6$, and
future surveys hold great promise to teach us even more.

This work was supported in part by NSF grants ACI AST 99-00877, AST
00-71019, AST 0098606, and PHY 0116590 and NASA ATP grants NAG5-12140
and NAG5-13292 and by the David and Lucille Packard Foundation
Fellowship for Science and Engineering.


\begin{thebibliography}{}

\bibitem[\protect\citeauthoryear{{Barkana}}{{Barkana}}{2002}]{barkana02-bub}
{Barkana} R.,  2002, New Astronomy, 7, 85

\bibitem[\protect\citeauthoryear{{Barkana}}{{Barkana}}{2004}]{barkana04-inf}
{Barkana} R.,  2004, MNRAS, 347, 59

\bibitem[\protect\citeauthoryear{{Barkana} \& {Loeb}}{{Barkana} \&
  {Loeb}}{2001}]{barkana01}
{Barkana} R.,  {Loeb} A.,  2001, Phys. Rep., 349, 125

\bibitem[\protect\citeauthoryear{{Barkana} \& {Loeb}}{{Barkana} \&
  {Loeb}}{2002}]{barkana02}
{Barkana} R.,  {Loeb} A.,  2002, ApJ, 578, 1

\bibitem[\protect\citeauthoryear{{Barkana} \& {Loeb}}{{Barkana} \&
  {Loeb}}{2003}]{barkana03}
{Barkana} R.,  {Loeb} A.,  2003, ApJ, submitted, (astro-ph/0310338)

\bibitem[\protect\citeauthoryear{{Barkana} \& {Loeb}}{{Barkana} \&
  {Loeb}}{2004}]{barkana04-grb}
{Barkana} R.,  {Loeb} A.,  2004, ApJ, 601, 64

\bibitem[\protect\citeauthoryear{{Barton} et~al.,}{{Barton}
  et~al.}{2004}]{barton04}
{Barton} E.~J.,  et~al., 2004, ApJ, 604, L1

\bibitem[\protect\citeauthoryear{{Becker} et~al.,}{{Becker}
  et~al.}{2001}]{becker01}
{Becker} R.~H.,  et~al., 2001, AJ, 122, 2850

\bibitem[\protect\citeauthoryear{{Bond}, {Cole}, {Efstathiou} \&
  {Kaiser}}{{Bond} et~al.}{1991}]{bond91}
{Bond} J.~R.,  {Cole} S.,  {Efstathiou} G.,    {Kaiser} N.,  1991, ApJ, 379,
  440

\bibitem[\protect\citeauthoryear{{Boyle} et~al.,}{{Boyle}
  et~al.}{2000}]{boyle00}
{Boyle} B.~J.,  et~al., 2000, MNRAS, 317, 1014

\bibitem[\protect\citeauthoryear{{Bromm}, {Kudritzki} \& {Loeb}}{{Bromm}
  et~al.}{2001}]{bromm-vms}
{Bromm} V.,  {Kudritzki} R.~P.,    {Loeb} A.,  2001, ApJ, 552, 464

\bibitem[\protect\citeauthoryear{{Cen}}{{Cen}}{2003a}]{cen03-qso}
{Cen} R.,  2003a, ApJ, 597, L13

\bibitem[\protect\citeauthoryear{{Cen}}{{Cen}}{2003b}]{cen03}
{Cen} R.,  2003b, ApJ, 591, L5

\bibitem[\protect\citeauthoryear{{Cen}, {Haiman} \& {Mesinger}}{{Cen}
  et~al.}{2004}]{cen04}
{Cen} R.,  {Haiman} Z.,    {Mesinger} A.,  2004, ApJ, submitted,
  (astro-ph/0403419)

\bibitem[\protect\citeauthoryear{{Ciardi}, {Stoehr} \& {White}}{{Ciardi}
  et~al.}{2003}]{ciardi03-sim}
{Ciardi} B.,  {Stoehr} F.,    {White} S.~D.~M.,  2003, MNRAS, 343, 1101

\bibitem[\protect\citeauthoryear{{Elvis} et~al.,}{{Elvis}
  et~al.}{1994}]{elvis94}
{Elvis} M.,  et~al., 1994, ApJS, 95, 1

\bibitem[\protect\citeauthoryear{{Fan} et~al.,}{{Fan}  et~al.}{2002}]{fan}
{Fan} X.,  et~al., 2002, AJ, 123, 1247

\bibitem[\protect\citeauthoryear{{Fan} et~al.,}{{Fan}  et~al.}{2003}]{fan03}
{Fan} X.,  et~al., 2003, AJ, 125, 1649

\bibitem[\protect\citeauthoryear{{Fan} et~al.,}{{Fan}  et~al.}{2004}]{fan04}
{Fan} X.,  et~al., 2004, AJ, in press, (astro-ph/0405138)

\bibitem[\protect\citeauthoryear{{Furlanetto}, {Zaldarriaga} \&
  {Hernquist}}{{Furlanetto} et~al.}{2004a}]{furl04b}
{Furlanetto} S.~R.,  {Zaldarriaga} M.,    {Hernquist} L.,  2004a, ApJ,
  submitted (astro-ph/0404112)

\bibitem[\protect\citeauthoryear{{Furlanetto}, {Zaldarriaga} \&
  {Hernquist}}{{Furlanetto} et~al.}{2004b}]{furl04a}
{Furlanetto} S.~R.,  {Zaldarriaga} M.,    {Hernquist} L.,  2004b, ApJ,
  submitted (astro-ph/0403697)

\bibitem[\protect\citeauthoryear{{Gnedin} \& {Prada}}{{Gnedin} \&
  {Prada}}{2004}]{gnedin04}
{Gnedin} N.~Y.,  {Prada} F.,  2004, ApJ, submitted, (astro-ph/0403345)

\bibitem[\protect\citeauthoryear{{Gunn} \& {Peterson}}{{Gunn} \&
  {Peterson}}{1965}]{gp}
{Gunn} J.~E.,  {Peterson} B.~A.,  1965, ApJ, 142, 1633

\bibitem[\protect\citeauthoryear{{Haiman}}{{Haiman}}{2002}]{haiman02}
{Haiman} Z.,  2002, ApJ, 576, L1

\bibitem[\protect\citeauthoryear{{Haiman} \& {Holder}}{{Haiman} \&
  {Holder}}{2003}]{haiman03}
{Haiman} Z.,  {Holder} G.~P.,  2003, ApJ, 595, 1

\bibitem[\protect\citeauthoryear{{Holder}, {Haiman}, {Kaplinghat} \&
  {Knox}}{{Holder} et~al.}{2003}]{holder03}
{Holder} G.~P.,  {Haiman} Z.,  {Kaplinghat} M.,    {Knox} L.,  2003, ApJ, 595,
  13

\bibitem[\protect\citeauthoryear{{Hu} et~al.,}{{Hu}  et~al.}{2002}]{hue02}
{Hu} E.~M.,  et~al., 2002, ApJ, 568, L75

\bibitem[\protect\citeauthoryear{{Hui} \& {Haiman}}{{Hui} \&
  {Haiman}}{2003}]{hui03}
{Hui} L.,  {Haiman} Z.,  2003, ApJ, 596, 9

\bibitem[\protect\citeauthoryear{{Kneib}, {Ellis}, {Santos} \&
  {Richard}}{{Kneib} et~al.}{2004}]{kneib04}
{Kneib} J.,  {Ellis} R.~S.,  {Santos} M.~R.,    {Richard} J.,  2004, ApJ, 607,
  697

\bibitem[\protect\citeauthoryear{{Kodaira} et~al.,}{{Kodaira}
  et~al.}{2003}]{kodaira03}
{Kodaira} K.,  et~al., 2003, PASJ, 55, L17

\bibitem[\protect\citeauthoryear{{Kogut} et~al.,}{{Kogut}
  et~al.}{2003}]{kogut03}
{Kogut} A.,  et~al., 2003, ApJS, 148, 161

\bibitem[\protect\citeauthoryear{{Lacey} \& {Cole}}{{Lacey} \&
  {Cole}}{1993}]{lacey}
{Lacey} C.,  {Cole} S.,  1993, MNRAS, 262, 627

\bibitem[\protect\citeauthoryear{{Leitherer} et~al.,}{{Leitherer}
  et~al.}{1999}]{leitherer}
{Leitherer} C.,  et~al., 1999, ApJS, 123, 3

\bibitem[\protect\citeauthoryear{{Loeb}, {Barkana} \& {Hernquist}}{{Loeb}
  et~al.}{2004}]{loeb04}
{Loeb} A.,  {Barkana} R.,    {Hernquist} L.,  2004, ApJ, submitted, \\
  (astro-ph/0403193)

\bibitem[\protect\citeauthoryear{{Madau}, {Meiksin} \& {Rees}}{{Madau}
  et~al.}{1997}]{mmr}
{Madau} P.,  {Meiksin} A.,    {Rees} M.~J.,  1997, ApJ, 475, 429

\bibitem[\protect\citeauthoryear{{Madau} \& {Rees}}{{Madau} \&
  {Rees}}{2000}]{madau00}
{Madau} P.,  {Rees} M.~J.,  2000, ApJ, 542, L69

\bibitem[\protect\citeauthoryear{{Miralda-Escud{\' e}}, {Haehnelt} \&
  {Rees}}{{Miralda-Escud{\' e}} et~al.}{2000}]{miralda00}
{Miralda-Escud{\' e}} J.,  {Haehnelt} M.,    {Rees} M.~J.,  2000, ApJ, 530, 1

\bibitem[\protect\citeauthoryear{{Miralda-Escude}}{{Miralda-Escude}}{1998}]{mi%
resc98}
{Miralda-Escude} J.,  1998, ApJ, 501, 15

\bibitem[\protect\citeauthoryear{{Oh}}{{Oh}}{2001}]{oh01b}
{Oh} S.~P.,  2001, ApJ, 553, 499

\bibitem[\protect\citeauthoryear{{Partridge} \& {Peebles}}{{Partridge} \&
  {Peebles}}{1967}]{partridge67}
{Partridge} R.~B.,  {Peebles} P.~J.~E.,  1967, ApJ, 147, 868

\bibitem[\protect\citeauthoryear{{Pell{\' o}} et~al.,}{{Pell{\' o}}
  et~al.}{2004}]{pello04}
{Pell{\' o}} R.,  et~al., 2004, A\&A, 416, L35

\bibitem[\protect\citeauthoryear{{Press} \& {Schechter}}{{Press} \&
  {Schechter}}{1974}]{press}
{Press} W.~H.,  {Schechter} P.,  1974, ApJ, 187, 425

\bibitem[\protect\citeauthoryear{{Rhoads} et~al.,}{{Rhoads}
  et~al.}{2004}]{rhoads04}
{Rhoads} J.,  et~al., 2004, ApJ, in press, (astro-ph/0403161)

\bibitem[\protect\citeauthoryear{{Rhoads} \& {Malhotra}}{{Rhoads} \&
  {Malhotra}}{2001}]{rhoads01}
{Rhoads} J.~E.,  {Malhotra} S.,  2001, ApJ, 563, L5

\bibitem[\protect\citeauthoryear{{Ricotti}, {Haehnelt}, {Pettini} \&
  {Rees}}{{Ricotti} et~al.}{2004}]{ricotti04}
{Ricotti} M.,  {Haehnelt} M.~G.,  {Pettini} M.,    {Rees} M.~J.,  2004, MNRAS,
  submitted, (astro-ph/0403327)

\bibitem[\protect\citeauthoryear{{Santos} et~al.,}{{Santos}
  et~al.}{2003}]{santos03}
{Santos} M.~G.,  et~al., 2003, ApJ, 598, 756

\bibitem[\protect\citeauthoryear{{Santos}}{{Santos}}{2004}]{santos04}
{Santos} M.~R.,  2004, MNRAS, 349, 1137

\bibitem[\protect\citeauthoryear{{Santos} et~al.,}{{Santos}
  et~al.}{2004}]{santos04-obs}
{Santos} M.~R.,  et~al., 2004, ApJ, 606, 683

\bibitem[\protect\citeauthoryear{{Scott} \& {Rees}}{{Scott} \&
  {Rees}}{1990}]{scott}
{Scott} D.,  {Rees} M.~J.,  1990, MNRAS, 247, 510

\bibitem[\protect\citeauthoryear{{Shapiro}, {Iliev} \& {Raga}}{{Shapiro}
  et~al.}{2004}]{shapiro04}
{Shapiro} P.~R.,  {Iliev} I.~T.,    {Raga} A.~C.,  2004, MNRAS, 348, 753

\bibitem[\protect\citeauthoryear{{Sheth}}{{Sheth}}{1998}]{sheth98}
{Sheth} R.~K.,  1998, MNRAS, 300, 1057

\bibitem[\protect\citeauthoryear{{Sheth} \& {Diaferio}}{{Sheth} \&
  {Diaferio}}{2001}]{sheth01-vel}
{Sheth} R.~K.,  {Diaferio} A.,  2001, MNRAS, 322, 901

\bibitem[\protect\citeauthoryear{{Sokasian}, {Abel} \& {Hernquist}}{{Sokasian}
  et~al.}{2002}]{sokasian02}
{Sokasian} A.,  {Abel} T.,    {Hernquist} L.,  2002, MNRAS, 332, 601

\bibitem[\protect\citeauthoryear{{Sokasian}, {Abel} \& {Hernquist}}{{Sokasian}
  et~al.}{2003}]{sokasian03a}
{Sokasian} A.,  {Abel} T.,    {Hernquist} L.,  2003, MNRAS, 340, 473

\bibitem[\protect\citeauthoryear{{Sokasian}, {Abel}, {Hernquist} \&
  {Springel}}{{Sokasian} et~al.}{2003}]{sokasian03}
{Sokasian} A.,  {Abel} T.,  {Hernquist} L.,    {Springel} V.,  2003, MNRAS,
  344, 607

\bibitem[\protect\citeauthoryear{{Sokasian} et~al.,}{{Sokasian}
  et~al.}{2003}]{sokasian03b}
{Sokasian} A.,  et~al., 2003, MNRAS, submitted, (astro-ph/0307451)

\bibitem[\protect\citeauthoryear{{Songaila}}{{Songaila}}{2004}]{songaila04}
{Songaila} A.,  2004, AJ, 127, 2598

\bibitem[\protect\citeauthoryear{{Spergel} et~al.,}{{Spergel}
  et~al.}{2003}]{spergel03}
{Spergel} D.~N.,  et~al., 2003, ApJS, 148, 175

\bibitem[\protect\citeauthoryear{{Springel} \& {Hernquist}}{{Springel} \&
  {Hernquist}}{2003}]{springel03}
{Springel} V.,  {Hernquist} L.,  2003, MNRAS, 339, 312

\bibitem[\protect\citeauthoryear{{Stanway} et~al.,}{{Stanway}
  et~al.}{2004a}]{stanway04-goods}
{Stanway} E.~R.,  et~al., 2004a, ApJ, 607, 704

\bibitem[\protect\citeauthoryear{{Stanway} et~al.,}{{Stanway}
  et~al.}{2004b}]{stanway04}
{Stanway} E.~R.,  et~al., 2004b, ApJ, 604, L13

\bibitem[\protect\citeauthoryear{{Theuns} et~al.,}{{Theuns}
  et~al.}{2002}]{theuns02-reion}
{Theuns} T.,  et~al., 2002, ApJ, 567, L103

\bibitem[\protect\citeauthoryear{{Venkatesan}, {Giroux} \&
  {Shull}}{{Venkatesan} et~al.}{2001}]{venkatesan01}
{Venkatesan} A.,  {Giroux} M.~L.,    {Shull} J.~M.,  2001, ApJ, 563, 1

\bibitem[\protect\citeauthoryear{{White}, {Becker}, {Fan} \& {Strauss}}{{White}
  et~al.}{2003}]{white03}
{White} R.~L.,  {Becker} R.~H.,  {Fan} X.,    {Strauss} M.~A.,  2003, AJ, 126,
  1

\bibitem[\protect\citeauthoryear{{Wyithe} \& {Loeb}}{{Wyithe} \&
  {Loeb}}{2003}]{wyithe03}
{Wyithe} J.~S.~B.,  {Loeb} A.,  2003, ApJ, 588, L69

\bibitem[\protect\citeauthoryear{{Wyithe} \& {Loeb}}{{Wyithe} \&
  {Loeb}}{2004a}]{wyithe04-lya}
{Wyithe} J.~S.~B.,  {Loeb} A.,  2004a, Science, submitted

\bibitem[\protect\citeauthoryear{{Wyithe} \& {Loeb}}{{Wyithe} \&
  {Loeb}}{2004b}]{wyithe04-prox}
{Wyithe} J.~S.~B.,  {Loeb} A.,  2004b, , 427, 815

\bibitem[\protect\citeauthoryear{{Zaldarriaga}}{{Zaldarriaga}}{1997}]{zal97}
{Zaldarriaga} M.,  1997, Phys Rev D, 55, 1822

\bibitem[\protect\citeauthoryear{{Zaldarriaga}, {Furlanetto} \&
  {Hernquist}}{{Zaldarriaga} et~al.}{2004}]{zald04}
{Zaldarriaga} M.,  {Furlanetto} S.~R.,    {Hernquist} L.,  2004, ApJ, in
  press, \\ (astro-ph/0311514)

\end{thebibliography}

\end{document}